\documentclass[aps,prc,nofootinbib,twocolumn,superscriptaddress]{revtex4}

\usepackage{graphicx}% Include figure files
\usepackage{dcolumn}% Align table columns on decimal point
\usepackage{bm}% bold math
\usepackage{lineno} % add linno

\usepackage[colorlinks=true,urlcolor=blue,linkcolor=blue,citecolor=blue]{hyperref}
\usepackage{etoolbox}
\usepackage{xcolor} % used to make text color 
\newcommand{\footremember}[2]{
   \footnote{#2}
    \newcounter{#1}
    \setcounter{#1}{\value{footnote}}
}
\newcommand{\footrecall}[1]{
    \footnotemark[\value{#1}]
} 

\begin{document}
%\linenumbers % Add line numbers to turn OFF \nolinenumbers
%\preprint{}

%\input{author-els.tex}

\title{Proton Form Factor Ratio, \boldmath $\mu_p G_E^p/G_M^p$ from Double Spin Asymmetry}



%%%%%%%
\author{A.~Liyanage\footnote{Corresponding author}}\affiliation{Hampton University, Hampton, VA}

%\author[JLab]{A.~Liyanage\corref{cor1}}
%\ead{anusha@jlab.org}

%%%%%%%%%%%%%%%%%%%%%%%%%%%%% full author list %%%%%%
\author{W.~Armstrong}\affiliation{Temple University, Philadelphia, PA}\affiliation{Argonne National Laboratory, Argonne, IL}
\author{H.~Kang}\affiliation{Seoul National University, Seoul, Korea}
\author{J.~Maxwell}\affiliation{University of Virginia, Charlottesville, VA}\affiliation{Thomas Jefferson National Accelerator Facility, Newport News, VA}
\author{J.~Mulholland}\affiliation{University of Virginia, Charlottesville, VA}
\author{L.~Ndukum}\affiliation{Mississippi State University, Starkville, MS}
\author{A.~Ahmidouch}\affiliation{North Carolina A\&M State University, Greensboro, NC}
\author{I.~Albayrak}\affiliation{Hampton University, Hampton, VA}
\author{A.~Asaturyan}\affiliation{Yerevan Physics Institute, Yerevan, Armenia}
\author{O.~Ates}\affiliation{Hampton University, Hampton, VA}
% ...

%%%%%%%%%%%%%%%%%%%%%%%%%%%%%%%%%%%%%%%%%%%
%\textcolor{red}{proton}

\date{\today}

\begin{abstract}
The ratio of the electric and magnetic form factor of the proton, $\mu_p G_E^p/G_M^p$, has been measured for elastic electron-proton scattering with polarized beam and target up to four-momentum transfer squared, $Q^2=5.66$ (GeV/c)$^2$ using the double spin asymmetry for target spin orientation aligned nearly perpendicular to the beam momentum direction.  

This measurement of $\mu_p G_E^p/G_M^p$ agrees with the $Q^2$ dependence of previous recoil polarization data and reconfirms the %dramatic
discrepancy at high $Q^2$ between the Rosenbluth and the polarization-transfer method with a different measurement technique and systematic uncertainties uncorrelated to those of the recoil-polarization measurements. The  form factor ratio at $Q^2$=2.06 (GeV/c)$^2$ has been measured as $\mu_p G_E^p/G_M^p = 0.720 \pm 0.176_{stat} \pm 0.039_{sys}$, which is in agreement with an earlier measurement with the polarized target technique at similar kinematics. The form factor ratio at $Q^2$=5.66 (GeV/c)$^2$ has been determined as $\mu_p G_E^p/G_M^p=0.244\pm0.353_{stat}\pm0.013_{sys}$, which represents the highest $Q^2$ reach with the double spin asymmetry with polarized target to date. 

\end{abstract}

\pacs{}

\maketitle

\section{Introduction}

The elastic form factors are fundamental properties of the nucleon, representing the effect of its structure on the response to electromagnetic probes such as electrons. Detailed knowledge of the nucleon form factors is critical for modeling of  the nucleus. Electron scattering is an excellent tool to probe deep inside nucleons and nuclei. In the one-photon exchange (Born) approximation, the structure of the proton or neutron is characterized by the electric and magnetic (Sachs) form factors, $G_E (Q^2 )$ and $G_M (Q^2 )$, which depend only on the four-momentum transfer squared, $Q^2$. At $Q^2=0$, the proton form factors are normalized to the charge $G_E^p(0)=1$ (in units of $e$) and the magnetic moment $G_M^p(0)=\mu_p=2.79$ (in units of nuclear magnetons). 

The Rosenbluth separation technique has been the first method to separate the squares of the proton form factors $G_E^p$ and $G_M^p$ by measuring the unpolarized elastic electron scattering cross sections at different angles and energies at fixed $Q^2$ \cite{rosenbluth1950}. In addition, the proton form factor ratio, $\mu_p G_E^p/G_M^p$ has been extracted from measurements of polarization components of the proton recoiling from the scattering of longitudinally polarized electrons \cite{157, 160}. In the ratio of polarization components, which is proportional to $G_E^p/G_M^p$, many of the experimental systematic
errors cancel.  

Measurement of the beam-target asymmetry using double polarization experiments with polarized target is a third technique to extract $\mu_p G_E^p/G_M^p$, which has not been conducted as often as Rosenbluth separation or recoil polarization experiments \cite{158,156}. For elastic scattering of polarized electrons from a polarized target, the beam-target double asymmetry, $A_p$ is directly related to the form factor ratio, $G_E^p/G_M^p$ as:
\begin {equation}
\label {asym}
A_p=\frac{-bR\sin\theta^*\cos\phi^*-a\cos\theta^*}{R^2+c},
\end {equation}
where $R=G_E^p/G_M^p$ with $R = 1/\mu_p$ at $Q^2=0$. 

The polar and azimuthal angles, $\theta^*$ and $\phi^*$ relative to the $z$ and $x$ axes, respectively, describe the orientation of the proton polarization vector relative to the direction of momentum transfer, $\vec{q} = \vec{p}_e - \vec{p}_{e'}$, where the $z$ axis points along $\vec{q}$, the $y$ axis perpendicular to the scattering plane defined by the electron three-momenta ($\vec{p}_e\times \vec{p}_{e'}$), and the $x$ axis so to form a right-handed coordinate frame. The quantities $a,b,c$ are kinematic factors given by $a=~2\tau\tan\frac{\theta_e}{2}\sqrt{1+\tau+(1+\tau)^2\tan^2\frac{\theta_e}{2}}$, $b=~2\tan\frac{\theta_e}{2}\sqrt{\tau(1+\tau)}$ and $c=\tau+2\tau(1+\tau)\tan^2\frac{\theta_e}{2}$ with $\tau=Q^2/(4M^2)$, where $\theta_e$ is the electron scattering angle and $M$ is the proton mass. 

\begin{figure}[b]
\includegraphics[width=\linewidth]{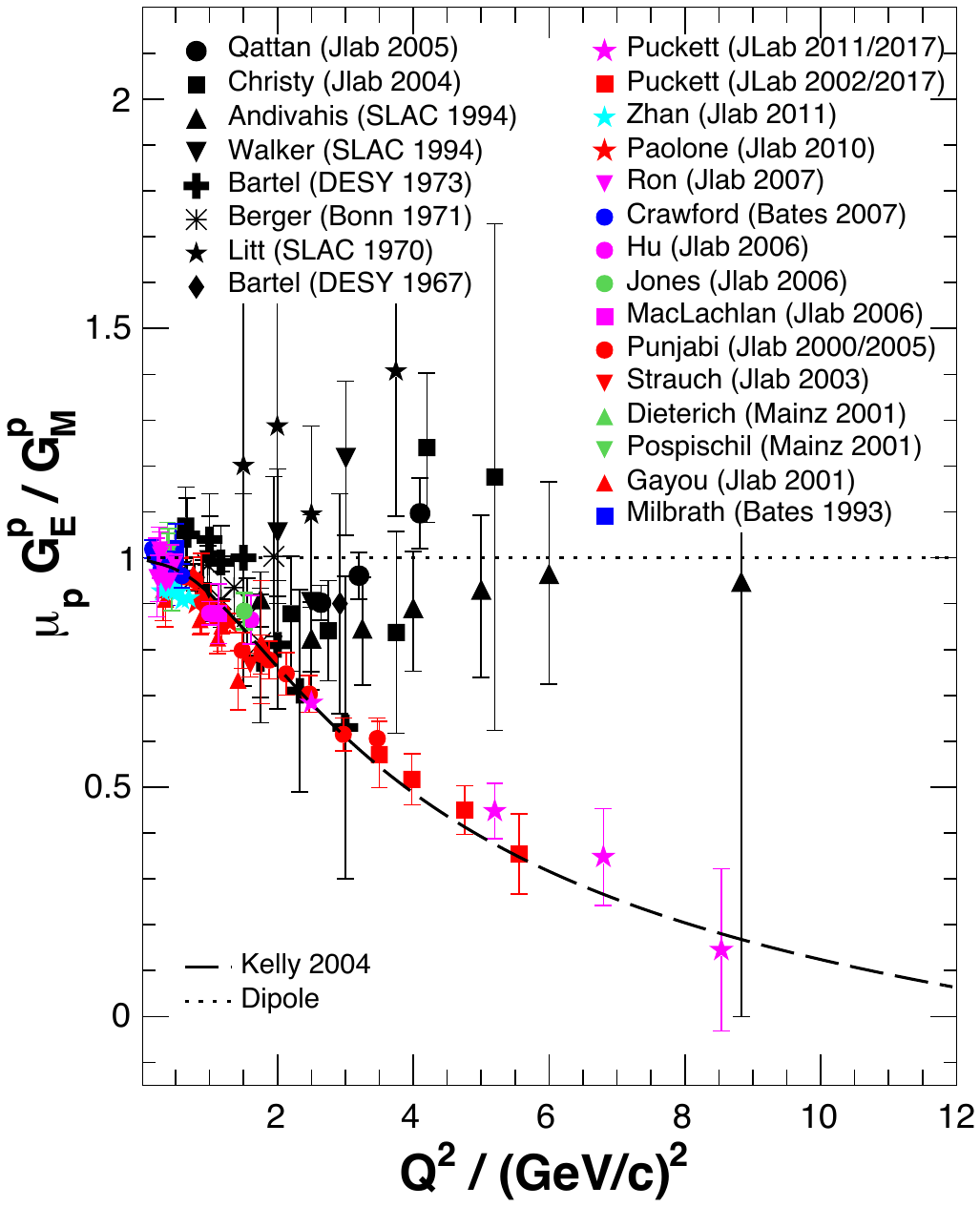}
\caption{Proton electric to magnetic form factor ratio from
  Rosenbluth-separated cross-sections, without TPE correction (\emph{black symbols}) \cite{26,
    27, 28, 29, 30, 31, 32, 33, 34, 35} and from double-polarization
  experiments (colored symbols) \cite{43, 44, 45, 46, 47, 48, 50,
    51, 52, 53, 54, 55, 56, 58, Puckett:2017flj, Puckett:2011xg, 59, 60}. The parametrization by
  Kelly \cite{99} is also shown.} 
\label{fig1}
\end{figure}

The world data of the proton form factor ratio, $\mu_pG^p_E/G^p_M$ from the Rosenbluth separation method \cite{26, 27, 28, 29, 30, 31, 32, 33, 34, 35}  are shown in Fig.~\ref{fig1} along with those obtained from double polarization experiments with recoil polarization \cite{43, 44, 45, 46, 47, 48, 50, 51, 52, 53, 54, 55, 56, 58, Puckett:2017flj, Puckett:2011xg} and polarized target \cite{59,60}. An almost linear fall-off of the polarization data can be seen compared to the nearly flat $Q^2$ dependence of $\mu_pG_E^p/G_M^p$ measured with the Rosenbluth technique. One possible solution that explains the difference between the polarized and unpolarized methods is two-photon exchange (TPE) \cite{97, 98, 96, 95, 94, 93, 92, 91, 90, 89}, which mostly affects the Rosenbluth data while the correction of the polarization data is small. It is also argued that effects other than TPE are responsible for the discrepancy \cite{Kondratyuk:2005kk, Kuraev:2007dn,
Pacetti:2016}. Several experiments have been conducted to validate the TPE hypothesis by probing the angular dependence of recoil polarization~\cite{43}, nonlinear dependence of unpolarized cross sections on $\epsilon$~\cite{163}, and by directly comparing $e^+p$ and $e^-p$ elastic scattering~\cite{154,161,162,155}. Evidence for TPE at $Q^2 < 2.5$ (GeV/c)$^2$ has been found to be smaller than expected, and more data are needed at high $Q^2$ to be conclusive~\cite{155}.

Having formally the equivalent sensitivity as the recoil polarization technique to the form factor ratio, the third technique,  beam-target asymmetry, is very well suited to verify the results of the recoil polarization technique. By measuring $\mu_pG_E^p/G_M^p$ and comparing it to the previous results, the discovery of any unknown or underestimated systematic errors in the previous polarization measurements is possible. The first such measurement was done by the experiment RSS at Jefferson Lab at $Q^2=1.5$~(GeV/c)$^2$ \cite{59}. Measurements of the form factor ratio at higher $Q^2$ values by a third technique, double-spin asymmetry, is an important consistency check on the results with the first two techniques, Rosenbluth separation and recoil polarization. In this work, the polarized target method has been applied at $Q^2$ = 2.06 and 5.66 (GeV/c)$^2$ as a by-product of the Spin Asymmetries of the Nucleon Experiment (SANE)~\cite{164}. 

Section II presents a description of the experimental setup. Section III discusses details of the data analysis method, including the elastic event selection, raw and physics asymmetry determinations, extraction of the proton form factor ratio, $\mu_pG^p_E/G^p_M$, and estimation of the systematic uncertainties. Section IV presents the final results of the experiment, which are discussed in Section V in light of the proton form factor ratio discrepancy. Section VI presents the conclusion with the impact of the measurement on the world database of the proton electromagnetic form factor ratio.

\section{Experimental Setup}

The experiment SANE (E07-003) is a single-arm inclusive-scattering experiment \cite{109, 144, 142, 143, 139, 140}. The goal of SANE was to measure the proton spin structure functions $g_1(x, Q^2)$ and $g_2(x, Q^2)$ at four-momentum transfer squared $2.5 < Q^2 < 6.5$ (GeV/c)$^2$ and values of the Bjorken scaling variable $0.3<x<0.8$, which is an extension of the kinematic coverage of experiment RSS performed in Hall C, Jefferson Lab \cite{179}.  
 
SANE measured the inclusive double spin asymmetries with the target spin aligned parallel and nearly perpendicular ($\approx$80$^{\circ}$) to the beam direction for longitudinally polarized electron scattering from a polarized target~\cite{SANE2017}. The experiment was carried out in experimental Hall ~C at Jefferson Lab from January to March, 2009. A subset of the data was used to measure the beam-target spin asymmetry from elastic electron-proton scattering for target spin orientation aligned nearly perpendicular to the beam momentum direction. Recoiled protons were detected by the High-Momentum Spectrometer (HMS) at $22.3^{\circ}$ and $22.0^{\circ}$, and central momenta of 3.58 and 4.17 GeV/c, for the two different beam energies 4.72 and 5.89 GeV, respectively. Scattered electrons were detected by the Big Electron Telescope Array (BETA) in coincidence with the protons in the HMS. In addition to that configuration, single-arm electron scattering data were also taken by detecting the elastically scattered electrons in the HMS at a central angle of 15.4$^{\circ}$ and a central momentum of 4.4 GeV/c for an electron beam energy of 5.89 GeV for both target spin configurations.

The Continuous Electron Beam Accelerator Facility (CEBAF) at the Thomas Jefferson National Accelerator Facility delivered longitudinally
polarized electron beams of up to 6 GeV with $\sim100$ \% duty factor simultaneously to the three experimental halls A, B, and C~\cite{100}. 
The CEBAF accelerator has recently been upgraded to 12 GeV with the addition of a fourth hall (D)~\cite{12GeV}. The Hall C arc dipole magnets
were used as a spectrometer to measure the energy of the electron beam as it entered the Hall. The beam polarization was measured with the Hall C M$\o$ller polarimeter \cite{104}. In addition to the standard Hall C beam-line instrumentation \cite{101, 102}, SANE required extra beamline equipment to accommodate a polarized target. Detailed description of the modifications to the standard Hall C beam line and the beam polarization can be found in \cite{SANENIM}. 

The primary apparatus for the elastic data was based on the superconducting High Momentum Spectrometer (HMS), which has a large solid angle and momentum acceptance, providing the capability of analyzing high momentum particles up to 7.4 GeV/c. A complete description of the HMS spectrometer and its performance during the SANE experiment in detail can be found in \cite{142}.  

In order to perform a coincidence experiment with the proton detected in HMS, the electron detector was required to have a large acceptance to match with the proton acceptance defined by the HMS collimator. The lead-glass electromagnetic calorimeter, BigCal as a part of BETA, provided the needed acceptance with sufficient energy and angular resolution for this coincidence electron determination \cite{SANENIM}.

As a double polarization experiment, SANE used a polarized proton target in the form of crystalized ammonia (NH$_3$). The protons in the  NH$_3$ molecules were polarized using Dynamic Nuclear Polarization (DNP)  \cite{136, 137, 138}. The target system consisted of a  target insert, a superconducting pair of Helmholtz magnets, a liquid helium evaporation refrigerator system and a Nuclear Magnetic Resonance (NMR) system. The target insert was roughly 2 m long, which provided room for four different containers of target materials, in 2.5 cm diameter cups. Two cups, called top and bottom, were filled with crystalized NH$_3$ beads, which were used as the proton targets. In addition to the crystalized ammonia, $^{12}$C and Polyethylene (CH$_2$) targets were also used for detector calibration purposes. The superconducting pair of Helmholtz magnets provided 5~T magnetic field in the target region. It can be rotated around the target in order to change the target field direction and hence the target polarization direction. More details on the operation of the target can be found in Ref.~\cite{109, SANENIM}. 

The beam-target asymmetry, $A_p$ shown in Eq.~(\ref{asym}), is maximal when the proton spin is aligned perpendicular to the four-momentum transfer direction. However, due to a constraint on the rotation of the Helmholtz magnets without blocking the BETA acceptance, the maximum spin direction one could reach was 80$^{\circ}$ relative to the beam direction, which was still acceptable enough for SANE's main physics \cite{109, 144, 143, 139, 140}.

\section{Data Analysis}
\subsection{Event Reconstruction}

The determination of the particle trajectory and momentum at the target using the HMS was done in two major steps. The first step was to find the trajectory, the positions and angles, $X_{fp}$ and $\theta_{fp}$ ($Y_{fp}$ and $\phi_{fp}$) in the dispersive (non-dispersive) direction at the detector focal plane using the two HMS drift chambers.  

The second step was to reconstruct the target quantities by mapping the focal plane coordinates to the target plane coordinates using a reconstruction matrix, which represents the HMS spectrometer optics based on a COSY model \cite{147}. This matrix was determined from previous data with the matrix that gives the correction due to the vertical target position being fixed to that determined from a COSY model. The reconstructed target quantities are $Y_{tar}$, $\phi_{tar}$, $\theta_{tar}$ and $\delta$, where $Y_{tar}$ is the horizontal position at the target plane perpendicular to the central spectrometer ray, $\phi_{tar}$ and $\theta_{tar}$ are the in-plane (non-dispersive) and out-of-plane (dispersive) scattering angles relative to the central ray. 
The HMS relative momentum parameter, $\delta = (P-P_0)/P_0$, where $P_0$ is the central momentum of the HMS, determines the momentum $P$ of 
the detected particle.

The presence of the target magnetic field affects the electron and proton trajectories. The standard matrix elements for $\delta$ and $\theta_{tar}$ take 
the vertical position of the beam at the target into account, hence the determinations of $\delta$ and of the out-of-plane angle, $\theta_{tar}$ are sensitive to a vertical beam position offset. The slow-raster system varied the vertical and horizontal position about the central beam location. The HMS optics matrix has been determined originally without the presence of a target magnetic field. Therefore, an additional particle transport through the target magnetic field has been added to the existing HMS particle-tracking algorithm to account for the additional particle deflection due to the target magnetic field. The treatment of this additional particle transport was developed in an iterative procedure. First, the particle track was reconstructed to the target from the focal plane quantities by the standard HMS reconstruction coefficients, assuming no target magnetic field but a certain vertical beam position. Using these target coordinates, the particle track was linearly propagated forward to the field-free region at 100~cm from the target center and then transported back to the target plane through the known target magnetic field, to determine the newly tracked vertical position. If a difference between the newly tracked vertical position at the target center and the assumed vertical position of the beam was observed then a new effective vertical position was assumed and the procedure was iterated until 
the difference between the tracked and assumed vertical positions became less than 1 mm~\cite{142}.  

\subsubsection{Corrections to HMS Event Reconstruction}

Comparisons of data and Monte Carlo simulation (MC) were used to determine the target vertical and horizontal position offsets relative to the beam center.  
In the MC, events were generated at assumed positions of the target and transported through the target magnetic field to an imaginary plane outside the field region. Then they were reconstructed back to the target using the standard HMS optics matrix. In the data, however, the events were reconstructed to the target positions using the same HMS optics matrix without the knowledge of the target offsets. The average target horizontal position offset, $X_{off}$=-0.15 mm, was determined by comparison of data to Monte Carlo simulation yields for the reconstructed horizontal position at the target, $Y_{tar}$ \cite{142}. 

The invariant mass, $W$ of the elastic $ep$ scattering can be written as
a function of the scattered electron momentum, $P$, angle, ${\theta_{e}}$
and beam energy, $E$ as  
\begin{equation}
\label{W}
W^2(P, \theta_e) = M^2 +2M(E-P) - 4 E P \sin^2{\theta_e}/2.
\end{equation}

\begin{figure}[b]
\centering
\begin{tabular}{cc}
\mbox{\includegraphics[width=\linewidth]{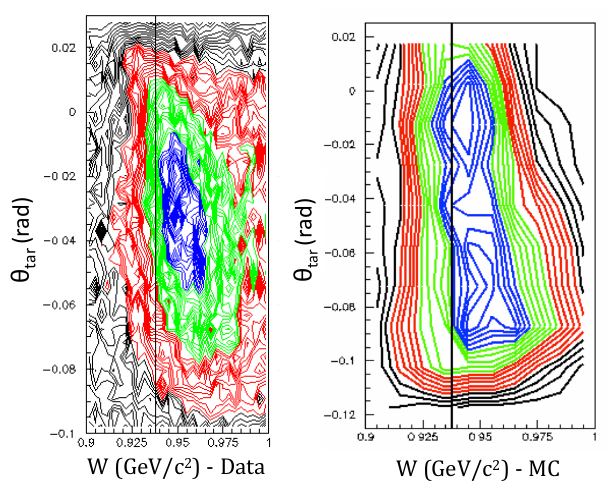}}
\end{tabular}
\caption{The correlation of $\theta_{tar}$ with $W$ for single-arm
  electron data  on HMS (left) and the same generated for MC (right).} 
\label{xoff_chk}                  
\end{figure}

In the single-arm data, the elastic peak in the $W$ spectrum was slightly correlated with $\theta_{tar}$ as in Fig.~\ref{xoff_chk} (left). Because both $\theta_{tar}$ and $\delta$ have first-order dependences on the vertical positions of the target in the reconstruction matrix element, the vertical beam position deviation from the target center, $Y_{off}$, can have effects on the reconstructed $\theta_{tar}$ as well as $\delta$ and hence $P$. This sensitivity caused the correlation of $\theta_{tar}$ with the invariant mass, $W$ as seen in Fig.~\ref{xoff_chk} (left). 

The same correlation can be reproduced by the Monte Carlo simulation by reconstructing the particle to a different vertical position than from where it was generated. The Monte Carlo generated correlation is shown in Fig.~\ref{xoff_chk} (right). Reproduction of the $\theta_{tar}$ vs $W$ correlation in MC generates confidence that the same correlation seen in the data is due to the reconstruction of the particle track to the incorrect vertical target position. Therefore, the average target vertical position offsets relative to the beam center were introduced and determined as +0.15  mm for the measured data by data-to-Monte Carlo simulation comparisons. This has been a suitable method to check the target vertical position offsets for the polarized target experiments.  

\subsubsection{Corrections to Coincidence Event Reconstruction}

The elastic events from the coincidence data were selected using both HMS and BigCal quantities. The horizontal (vertical) coordinate of the scattered electron at the front face of BigCal, $X_{BETA} (Y_{BETA})$ was measured using the shape of the energy distribution in the lead glass blocks. Assuming elastic kinematics, the proton momentum and angle measured by the HMS when combined with the beam energy can be used to calculate the scattered electron's kinematics. Using the predicted electron's momentum and angle, the electron can be tracked from the target through the magnetic field to the front face of BigCal to predict the horizontal ($X_{HMS}$) and vertical positions ($Y_{HMS}$). The differences between the measured and the calculated BETA quantities, ~$\Delta Y = (Y_{HMS}-Y_{BETA})$, and $\Delta X = (X_{HMS}-X_{BETA})$ was obtained and utilized to check the quality of the HMS-BETA coincidence data.

Based on energy and momentum conservation for electron-proton elastic scattering,  the recoil proton momentum, $P_p(\theta_p)$ could
be calculated from $\theta_p$, as
\begin{equation} 
\label{pp}
P_p(\theta_p)=\frac{2 M E(E + M) \cos
  {\theta_p}}{M^2+2 M E + E^2 \sin^2{\theta_p}}. 
\end{equation}
The residual difference between the proton momentum detected by HMS, $P_p$ and the proton momentum calculated by the recoiled proton angle, $P_p(\theta_p)$, expressed as a percentage of the HMS central momentum, $P_0$, is given as  
\begin{equation}
\label{delta_p}
{\Delta_p=\frac{P_p - P_p(\theta_p)}{P_0}.}
\end{equation}

\begin{figure}[b]
\centering
\begin{tabular}{cc}
\includegraphics[width=\linewidth]{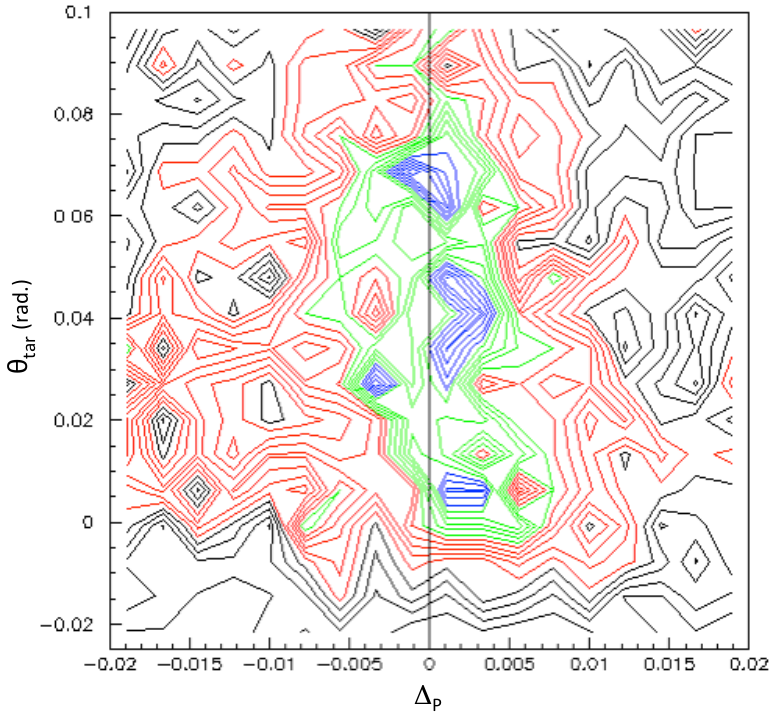}
\end{tabular}
\begin{tabular}{cc}
\includegraphics[width=\linewidth]{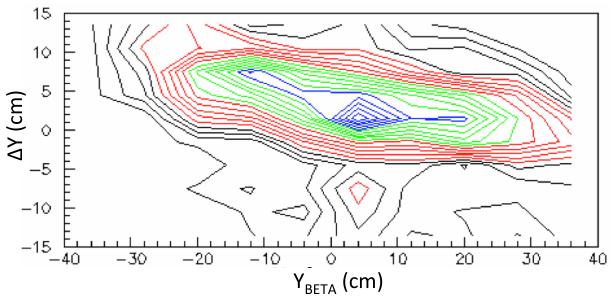}
\end{tabular}    
\caption{The correlation of the HMS quantities,  $\theta_{tar}$ vs
  $\Delta_p$ (\emph{Top}) and the correlation of the BETA quantities,
  $\Delta Y$ vs $Y_{BETA}$ (\emph{Bottom}) for the coincidence data.} 
\label{coin_corr}                  
\end{figure}

Correlations of the HMS quantities $\theta_{tar}$ vs $\Delta_p$, and the BETA quantities $\Delta$Y vs $Y_{BETA}$, were observed in the coincidence data, as seen in Fig.~\ref{coin_corr}. Since all of these correlations are related to the vertical position or angle, a correction of out-of-plane angle due to the target magnetic field was found to be the best explanation. Subsequently, all these correlations were corrected by applying an out-of-plane angle dependence to the magnet field used in the reconstruction of particle tracks from the target to the BigCal front face. This correction changed the particle's reconstructed momentum and, therefore, the reconstructed vertical position, which eliminated the above correlations. 
 
\subsection{Elastic Event Selection}
Single-arm electrons were identified in HMS with PID and momentum acceptance cuts. The Cherenkov and the lead glass calorimeter in HMS were used to discriminate $e^-$ from $\pi^-$, requiring the number of photoelectrons seen by the Cherenkov counter $N_{cer} > 2$ (Cherenkov cut) and the relative energy deposited in the lead glass calorimeter, $E_{cal}/P > 0.7$ (calorimeter cut), where $P$ is the reconstructed electron momentum in the HMS spectrometer \cite{142}.  

\begin{figure}[b]
\centering
\includegraphics[width=\linewidth]{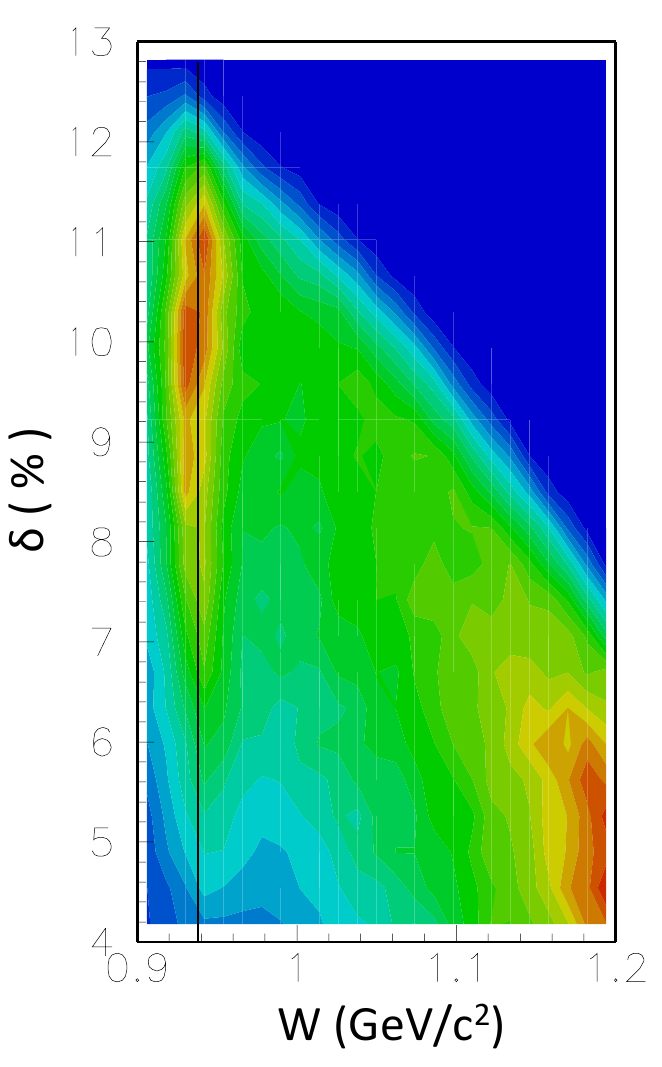}
\caption{
  The relative momentum $\delta$ for the single-arm elastic electron data as a function of invariant mass, $W$.} 
\label{dccalib}                  
\end{figure}

Figure~\ref{dccalib} shows the relative momentum, $\delta$, for the single-arm electron data as a function of invariant mass, $W$. The nominal momentum acceptance is given by $-8\% < \delta < 10\%$, which is usually applied as a fiducial cut in addition to the PID cuts. This eliminates events that are outside of
the nominal spectrometer acceptance, but end up in the detectors after multiple scattering in the magnets or exit windows. Because a significant number of elastic events populated the region of larger $\delta$ values, where the reconstruction matrix elements are not well known, these data were analyzed individually so that the systematic effect from the HMS reconstruction matrix could be determined separately. Therefore, two $\delta$ regions, $-8\% < \delta < 10\%$ and $10\% < \delta < 12\%$, were used separately in addition to the PID cuts to extract the elastic events. About $\sim40\%$ of extra elastic events were obtained by using the higher $\delta$ region.  

\begin{figure}[t]
\centering
\includegraphics[width=\linewidth]{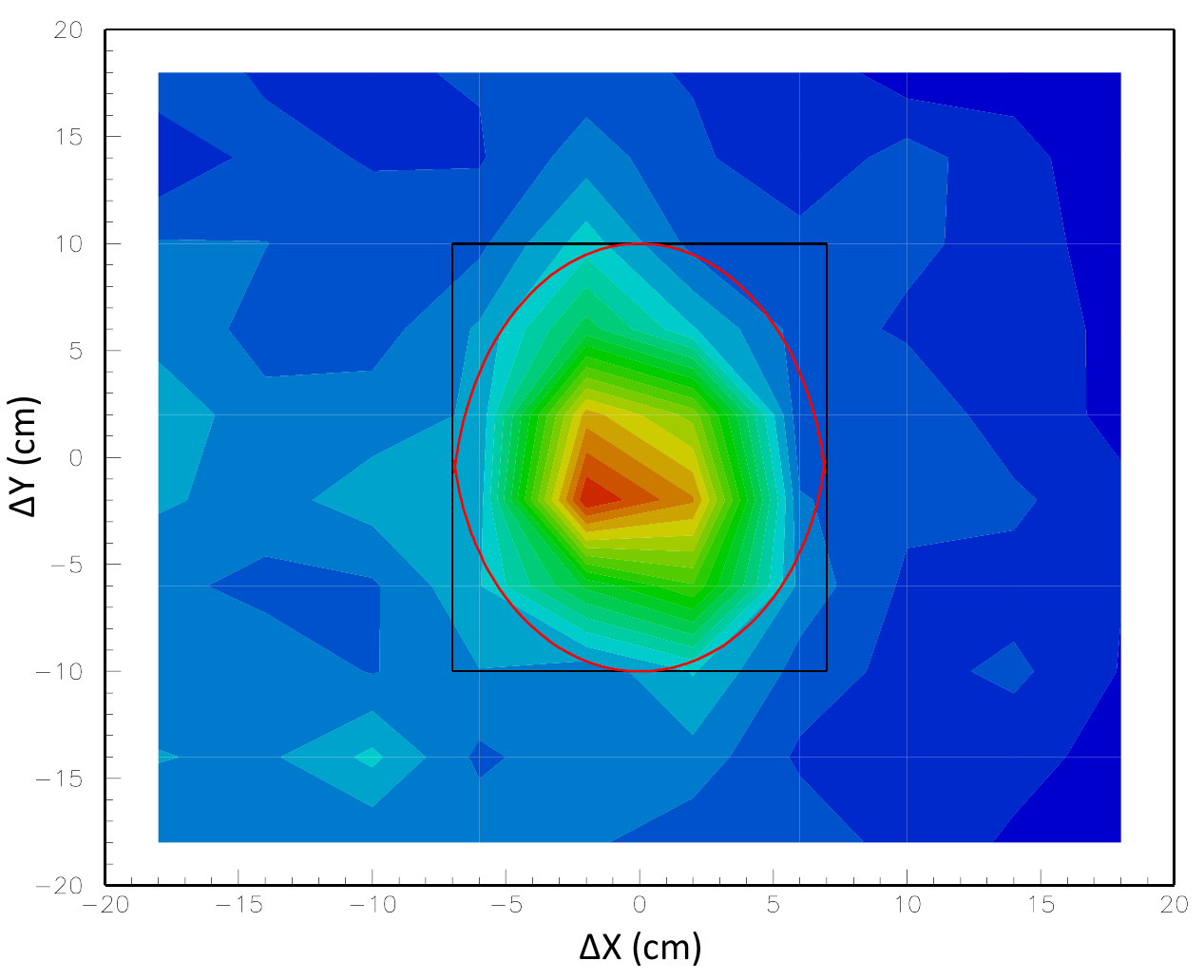}
\caption{A square cut with $\Delta X = \pm$ 7 cm and $\Delta Y = \pm$ 10 cm (black square) and the elliptical cut (red) with $(X_{cut},Y_{cut})=(7,10)$ cm were applied to the $\Delta Y$ vs $\Delta X$ distributions at $Q^2 = 6.19$ (GeV/c)$^2$.}
\label{xydiff}
\end{figure}

Both  HMS and BigCal quantities were used to select the elastic events from the coincidence data. The differences between the measured and the calculated BETA quantities, $\Delta Y$, and $\Delta X$ are shown in Fig.~\ref{xydiff}. A square cut applied with $\Delta X = \pm$ 7 cm and $\Delta Y = \pm$ 10 cm as in Fig.~\ref{xydiff} (black square) to reduce the background. However, an elliptic cut applied to the differences, $\Delta Y$, and $\Delta X$, 
\[
\left(\frac{\Delta X}{X_{cut}}\right)^2+\left(\frac{\Delta
      Y}{Y_{cut}}\right)^2 \le 1, \] 
with $X_{cut}$ and $Y_{cut}$ representing the half axes, reduces the backgrounds most effectively, as illustrated in Fig.~\ref{xydiff} (red circle).  Here, $(X_{cut}, Y_{cut}) = (7, 10)$ cm. 

\begin{figure}[hb]
\centering
\includegraphics[width=\linewidth]{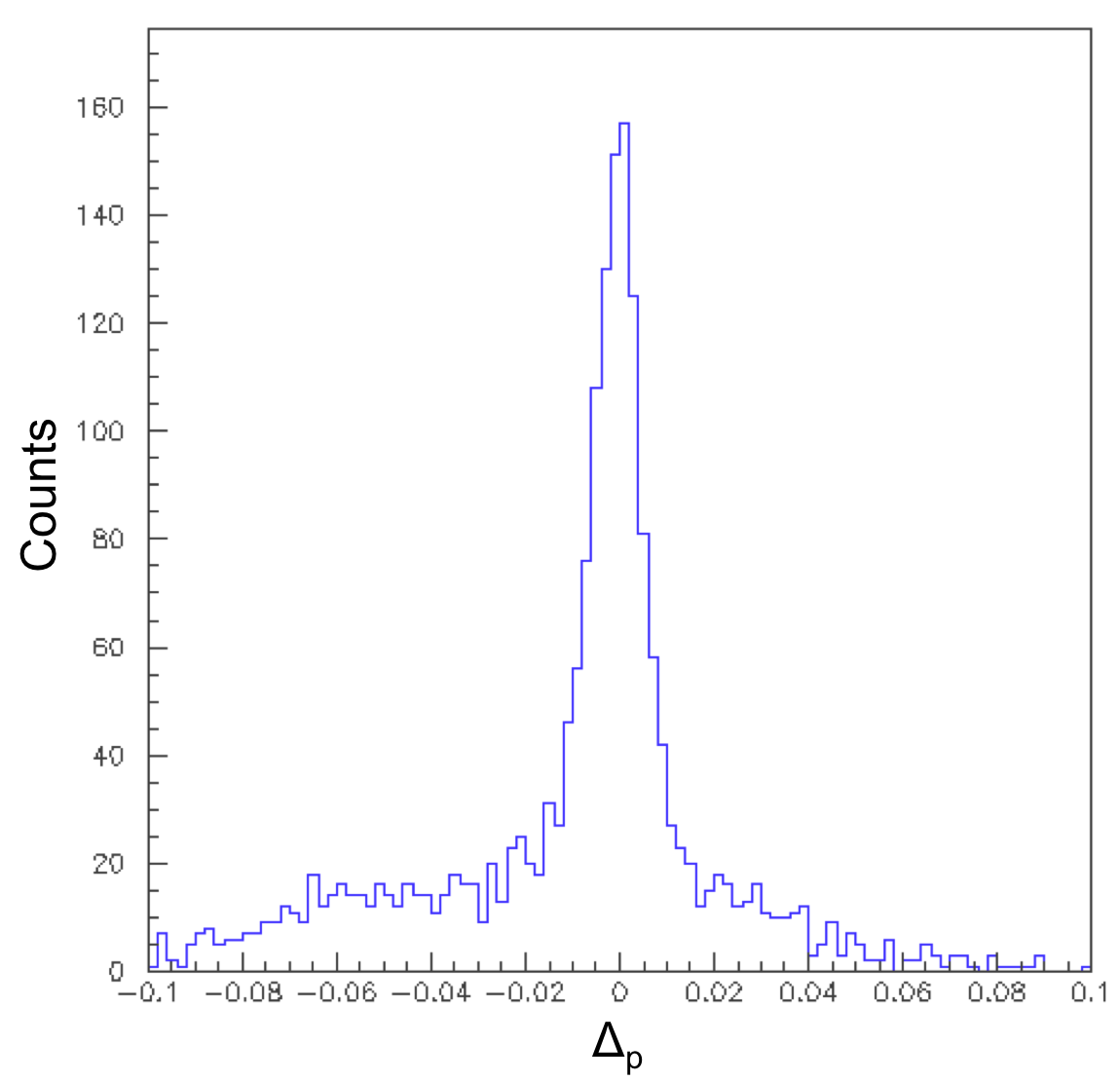}
\caption{The $\Delta_p$ spectrum of all coincidence events at $Q^2=6.19$ (GeV/c)$^2$ after applying the elliptical cut.} 
\label{Delta_p}
\end{figure}

The variance of $\Delta_p$, Eq. \ref{delta_p}, was found to be 0.7$\%$. A $\pm 3\sigma$ cut around the central peak of $\Delta_p$ was chosen for further background suppression for the coincidence data. The spectrum of $\Delta_p$ is shown in Fig.~\ref{Delta_p}.

\subsection{Raw/ Physics Asymmetries} %done up to here
The measured double polarization raw asymmetries of the extracted elastic events were formed by  
%scattering of polarized electrons from a transversely polarized
%target, the beam-target asymmetry, 
\begin {equation}
%\begin {aligned}
\label {rasym}
A_{raw}=\frac{N^+-N^-}{N^++N^- }, 
%\end {aligned}
\end {equation}                            
where $N^+$ and $N^-$ are the raw elastic yields normalized by the dead time corrected charge. They are defined by $N^+=N^{\uparrow\uparrow} +  N^{\downarrow\downarrow}$ and $N^-=N^{\uparrow\downarrow} +  N^{\downarrow\uparrow}$, where the first index refers to the beam helicity and the second index refers to the target polarization.  

The physics asymmetry,
\begin {equation}
%\begin {aligned}
\label {pasym}
A_p=\frac{A_{raw}}{P_B P_Tf} +N_c,
%\end {aligned}
\end {equation}                 
was obtained by dividing $A_{raw}$ by target and beam polarizations, $P_T$ and $P_B$, and the dilution factor, $f$.  

The dilution factor is the ratio of the yields of scattering off free protons to those from the entire target. The $N_c$ term is a correction to the measured raw asymmetry to account for the quasi-elastic scattering contribution from polarized $^{14}N$. For SANE, $N_c$ is larger and of opposite sign than for RSS \cite{59} because SANE used $^{14}N$ instead of $^{15}N$ in RSS. Therefore, the $N_c$ term for SANE is found to be $0.98$. 

The ratio of the volume taken by the ammonia crystals to the entire target cup volume is known as the \emph{packing fraction}, which was determined by normalizing the measured data with the simulated yields. The different packing fractions give rise to different target material contributions inside the target cup. Both target cups were used during the data taking. The packing fractions were determined for the top and bottom targets as (55$\pm$5$)\%$ and (60$\pm$5$)\%$ respectively. More details about the packing fraction determination can be found in Ref.~\cite{140, SANENIM}. 

\subsubsection{Determination of $f$ and A$_p$ for The Single-Arm Data}

\begin{figure*}
\centering
\begin{minipage}[]{0.49\textwidth}
\includegraphics[width=\linewidth]{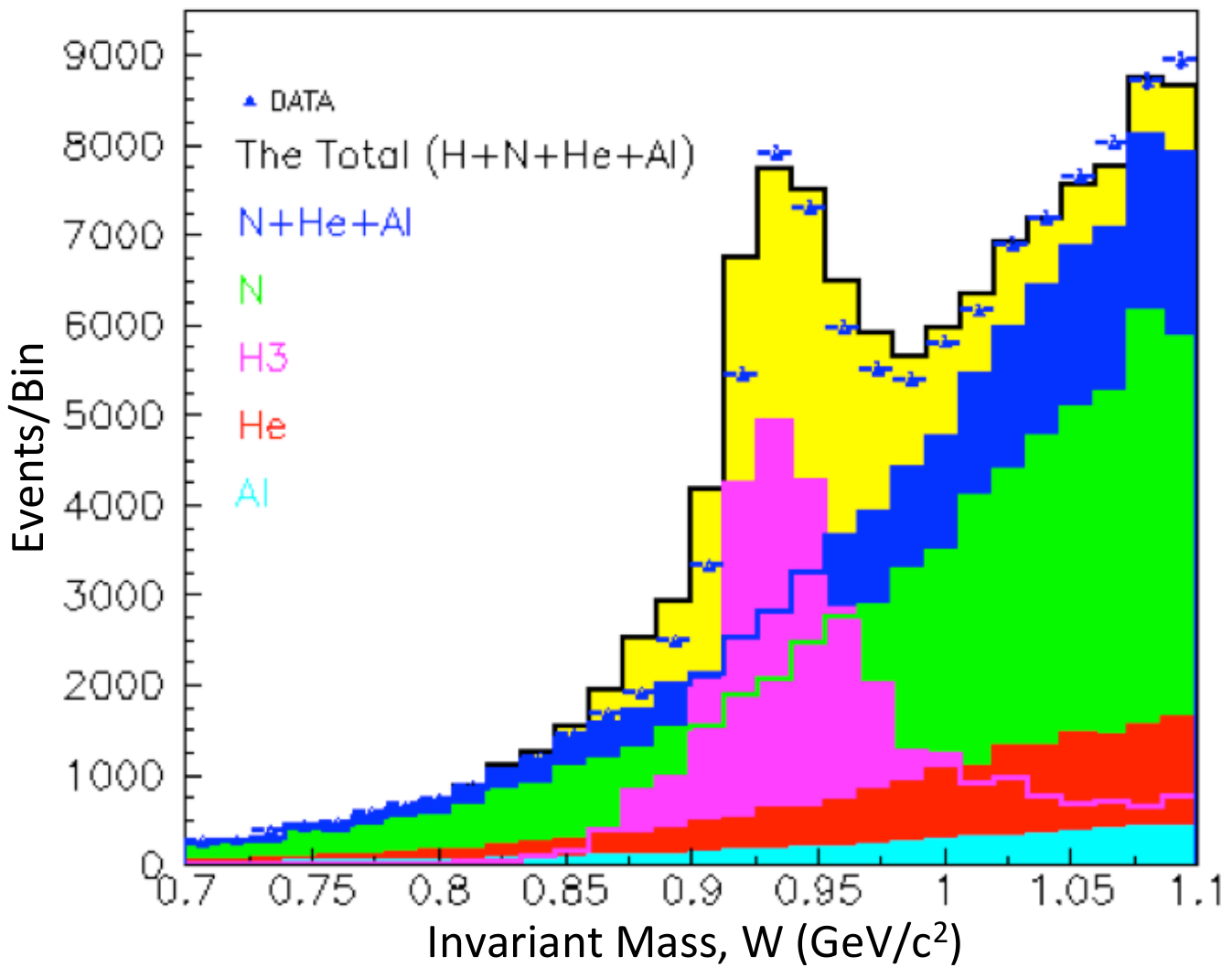}\\
\includegraphics[width=\linewidth]{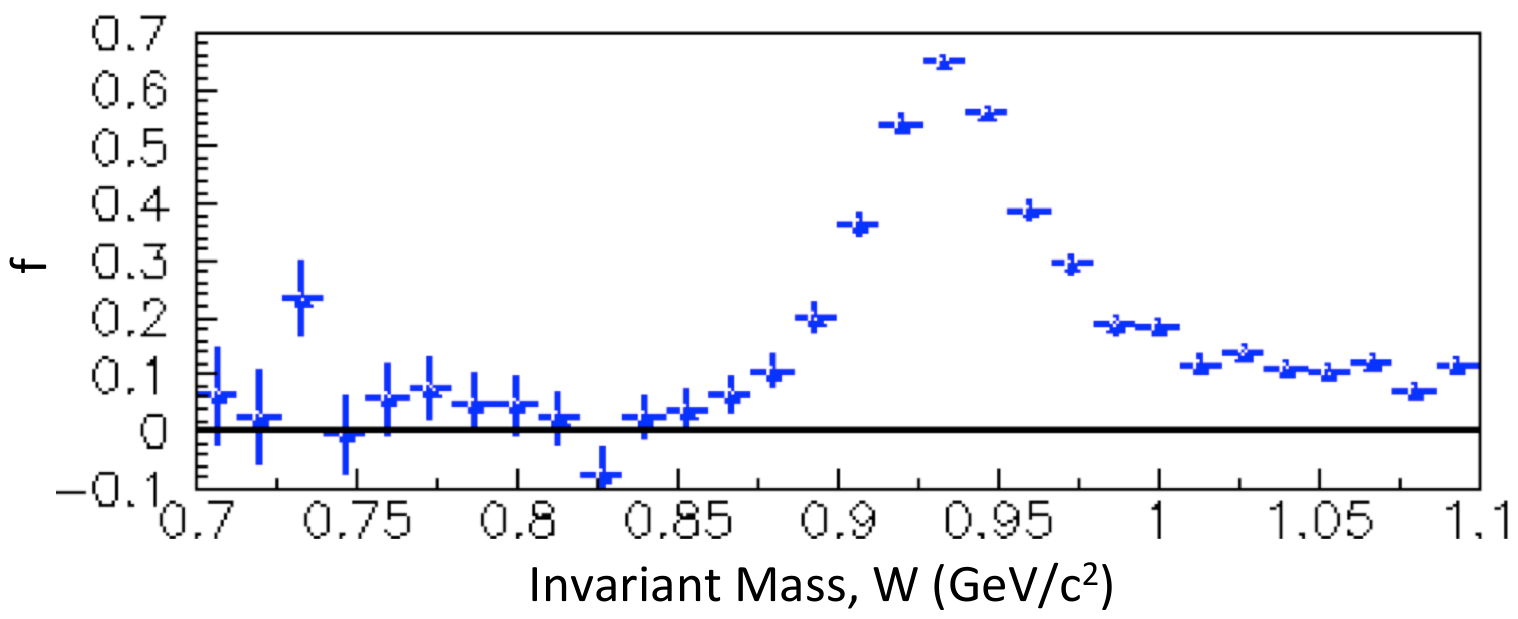}\\
\includegraphics[width=\linewidth]{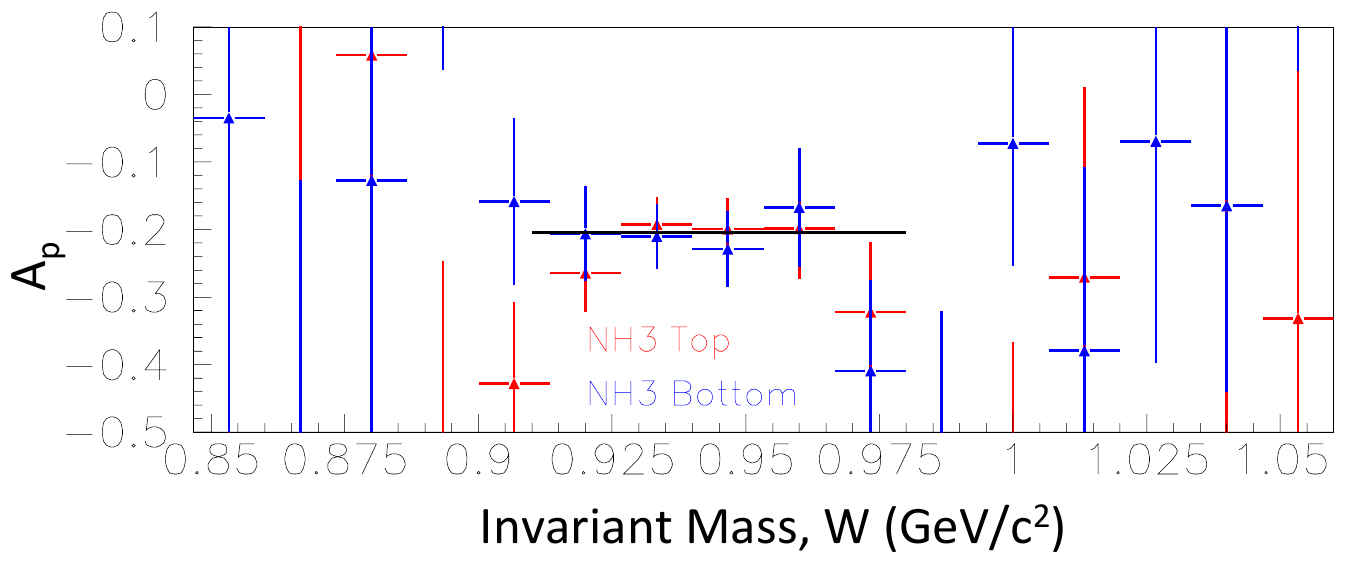}
\end{minipage}
\begin{minipage}[]{0.49\textwidth}
\includegraphics[width=\linewidth]{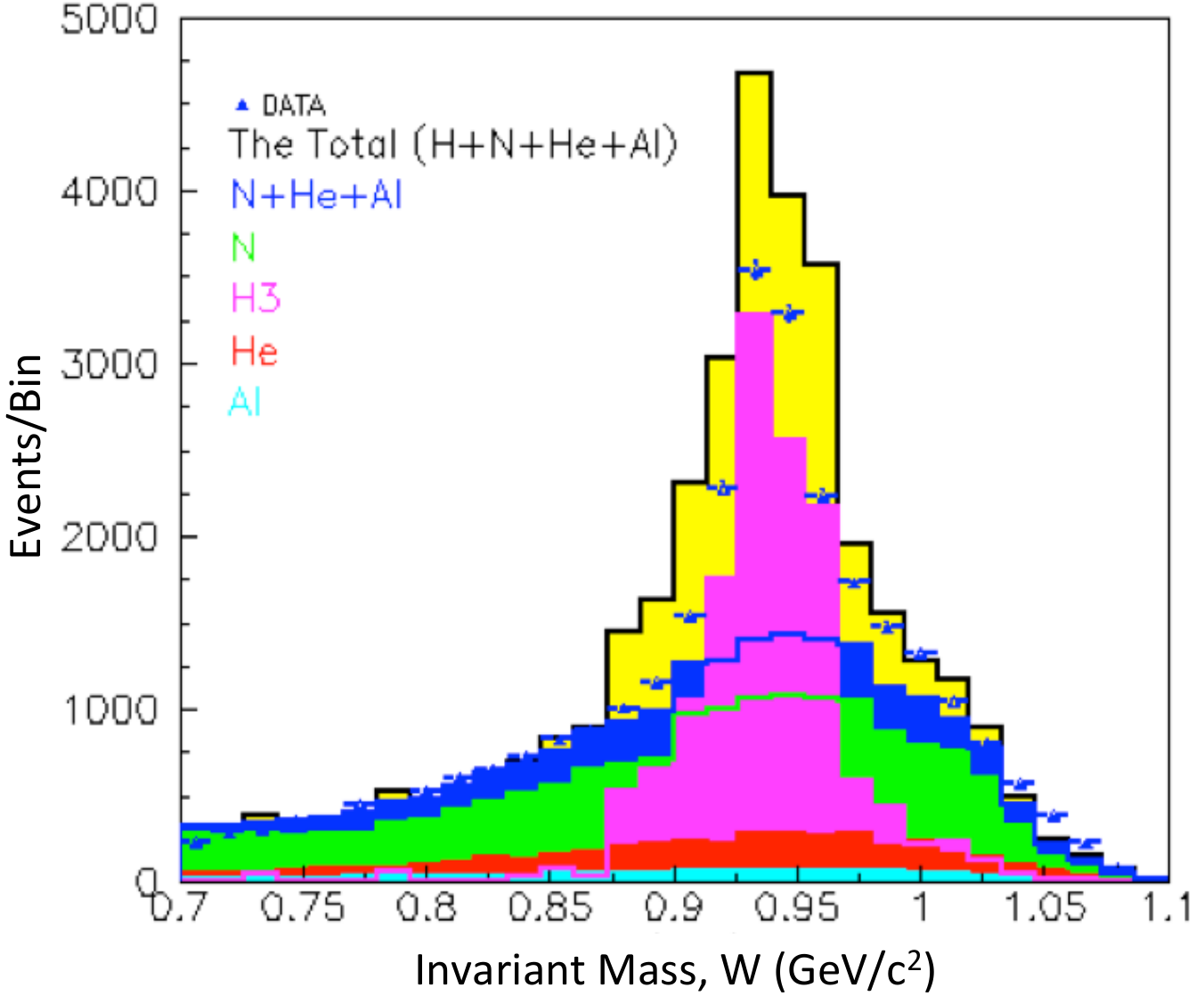}\\
\includegraphics[width=\linewidth]{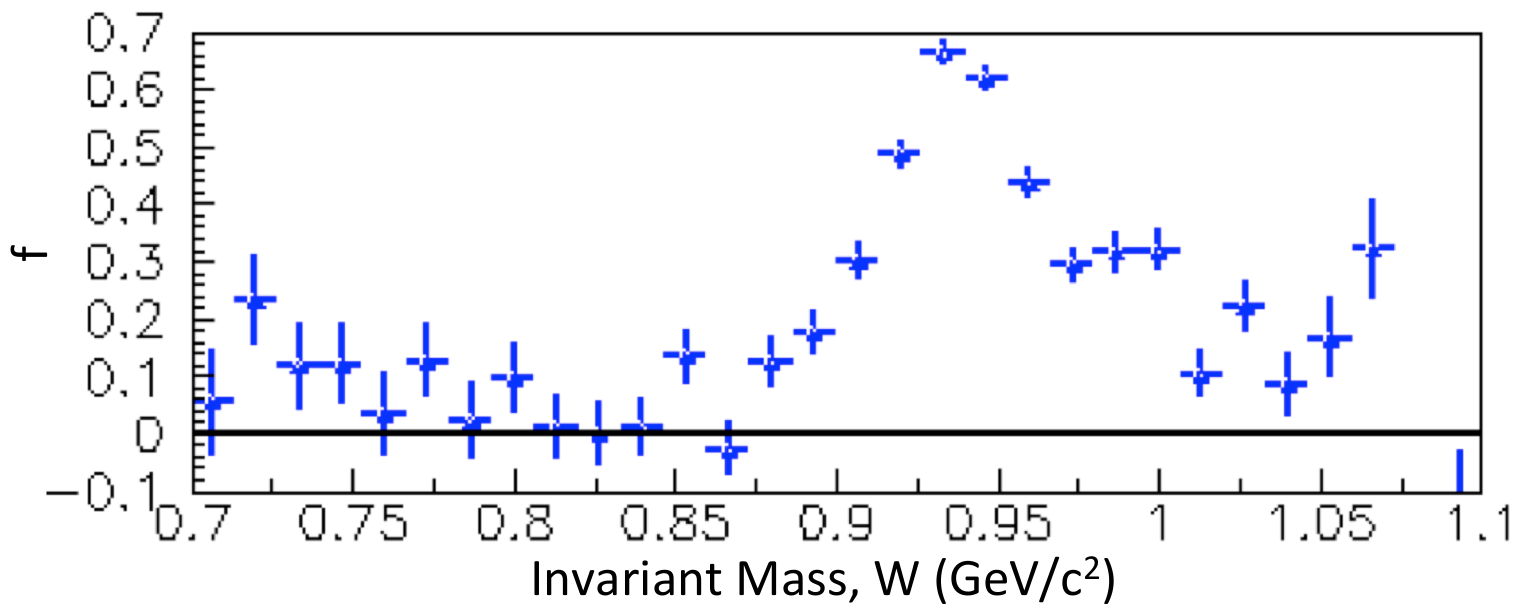}\\
\includegraphics[width=\linewidth]{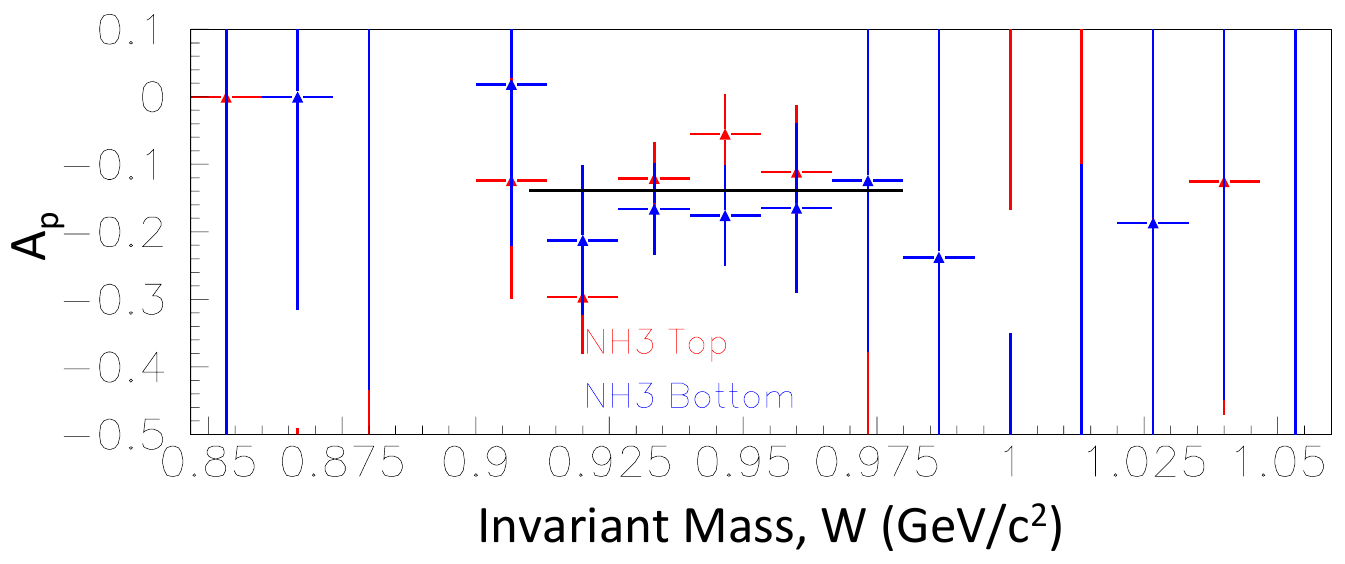}
\end{minipage}
%\begin{minipage}[]{0.42\textwidth}
%\includegraphics[width=\linewidth]{replot4paper_apsingle}
%\end{minipage}
\caption{
  Yields, dilution factor, and physics asymmetries as a function of $W$ for $-8\% < \delta < 10\%$ (\emph{left column}) and $10\% < \delta < 12\%$ (\emph{right column}). \emph{Top row}: The simulated target contributions at the elastic peak compared to the data as a function of $W$ for the top target. Different colors show different target material contributions to the yield. No normalization factors were used to match the MC to the data. \emph{Middle row}: The dilution factors inferred from simulated yields as a function of $W$ for the top target. \emph{Bottom row}: The resulting physics asymmetries for the top and bottom targets as a function of $W$.}
\label{trg_cont}                  
\end{figure*}

The dilution factor, $f$, represents the fraction of polarizable material in the beam from which electrons can scatter. The SANE target was immersed in a liquid He bath. Hence, electron scattering can occur from all the material inside the target cup, as well as from all the material in the beam path toward the target cup. The material consisted  of hydrogen (H), nitrogen (N), helium (He) and aluminum (Al). A Monte Carlo simulation was used to estimate these backgrounds in order to determine the dilution factor. The weighted amount of target materials inside each target cup was calculated, taking into account the packing fraction. The scattering yields due to H, N, He and Al were simulated using their individual cross sections \cite{Bosted:2007xd} and compared with
the single-arm elastic data to estimate the backgrounds. The simulated target contributions for the top target for the two different $\delta$ regions are shown in Fig.~\ref{trg_cont} (top row). In Fig.~\ref{trg_cont} (top right), the MC tail serves to estimate the background most accurately. However, the acceptance in the high-momentum bin is not well known, hence the peak yield deviates from the data.  Nevertheless, the spin asymmetry should still be accurate as it is mostly independent of the acceptance.

The dilution factors were calculated for both top and bottom targets by taking the ratio of the difference between the total raw yields and the Monte Carlo background radiated yields (N+He+Al) to the total raw yield,    
\begin{equation}
\label{dfsin}
f=\frac{Y_{data}-Y_{MC}}{Y_{data}},
\end{equation}
where $Y_{data} = N_+ + N_-$ is the total raw yield of the measured data and $Y_{MC}$ is the total Monte Carlo background yield from N, He, and Al. The obtained dilution factors are shown in Fig.~\ref{trg_cont} (middle row) for the top target for two different $\delta$ regions. The dilution factor is the largest in the elastic region where $0.91 < W < 0.97$ GeV/c$^2$. 

The physics asymmetry, $A_p$, was evaluated for the selected elastic events using Eq.~(\ref{pasym}) for average values of $P_B=(73 \pm 1.5)\% $, $P_T=(70 \pm 5.0)\%$, and by normalizing with the dilution factor, $f$. Figure~\ref{trg_cont} (\emph{bottom row}) shows the physics asymmetries for the top and bottom targets and for the two different $\delta$ regions, as a function of $W$. The physics asymmetries were constant in the elastic region of $0.91 < W < 0.97$ GeV/c$^2$, where the dilution factor is the largest, which supports that the functional dependence of $f$ on $W$ as in Fig.~\ref{trg_cont} (middle) is accurate. The average physics asymmetries and uncertainties of this constant region were determined for both targets and $\delta$ regions using an error-weighted mean of the $W$ bins in the interval of 0.91\textless W\textless0.97 GeV/c$^2$. The weighted average $A_p$ was obtained for each $\delta$ region by combining the average physics asymmetries from both top and bottom targets. The weighted average asymmetry results are shown in Fig.~\ref{physasym} (\emph{left}), and are listed in Tab.~\ref{tabasym} (\emph{left half}). 
 
\subsubsection{Determination of $f$ and A$_p$ for The Coincidence Data}

\begin{figure}[b]
\centering
\includegraphics[width=\linewidth]{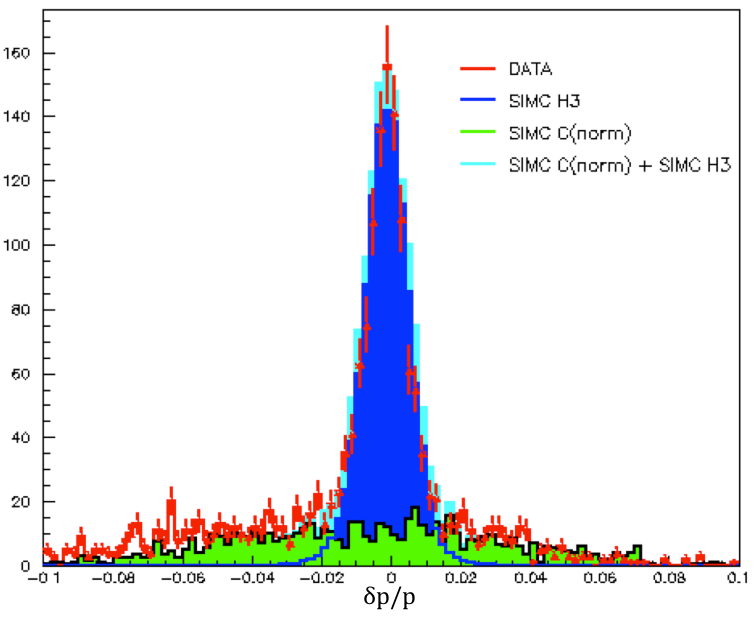}
\caption{The normalized carbon background (green) and H (blue) comparison to the coincidence data (red), which the background is subtracted by applying the elliptical cut as in Fig. \ref{xydiff}, for the beam energy 5.895 GeV.} 
\label{apcoin}                  
\end{figure}

For the coincidence data, the Monte Carlo simulation was generated using known H elastic cross sections \cite{Bosted:1994tm} and a model of quasi-free elastic scattering for carbon. The actual background material of helium, nitrogen and aluminum would have a similar background shape to carbon
and so carbon was used to represent the background shape and it was normalized to match the data in the region outside of the elastic peak. The region of $0.03 < \Delta_p < 0.08$, where the data and the background distributions match each other, was used to determine the normalization factor and hence the background shape under the elastic peak. A comparison between the measured data and the simulated yields is shown in Fig.~\ref{apcoin}.  The elastic data were extracted by applying the elliptic cut on $\Delta Y$ vs $\Delta X$ as in Fig.~\ref{xydiff}, which suppresses the background most effectively.
% and therefore, the data eliminate radiative tails.

Because of low statistics, the dilution factor for the coincidence data was not calculated as a function of $W$ (or $\Delta_p$), as done for elastic single-arm data. Instead, the average dilution factor was determined by an integration method using the normalized carbon MC yields and the measured data yields under the elastic peak  in the interval of  $| \Delta_p | \textless $0.02 $(3 \sigma)$ and then by using Eq.~(\ref{dfsin}). The procedure was done separately for
both beam energies, 5.895 GeV and 4.725 GeV. The average dilution factors based on the integration method for the top and bottom targets for the beam energy of 5.895 GeV were determined as $f = 0.785 \pm 0.039$ and $0.830 \pm 0.042$, respectively. Only the bottom target was used for 4.725 GeV and the dilution factor was determined as $f = 0.816 \pm 0.041$. 

The weighted average physics asymmetry and uncertainty between the top and bottom targets for the beam energy of 5.895 GeV were obtained as $A_p = 0.083 \pm 0.074$, while that for the beam energy of 4.725 GeV resulted in $A_p = 0.248 \pm 0.138$.

\begin{figure}[b]
\centering
\begin{tabular}{cc}
%\mbox{\includegraphics[width=0.44\linewidth]{replot4paper_apsingletb_onlyavg}}
%\mbox{\includegraphics[width=0.48\linewidth]{simc_ap_onlyavg}}
\includegraphics[width=\linewidth]{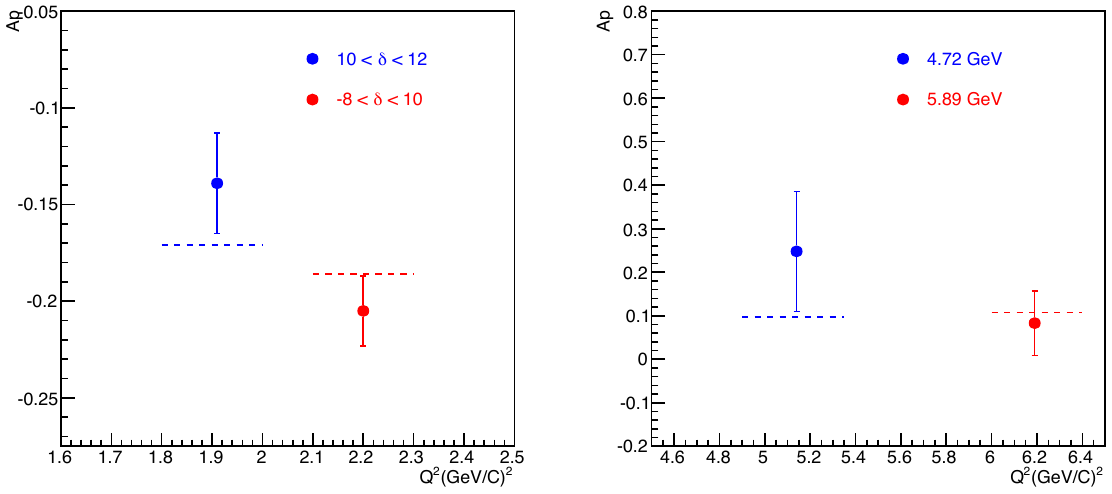}
\end{tabular}
\caption{
(\emph{Left}): The weighted average physics asymmetries for two different $\delta$ regions as a function of $Q^2$. The expected physics asymmetries from the known form factor ratio for each $Q^2$ by Kelly's form factor parametrization \cite{99} are also shown by dashed lines separately for the two different $\delta$ regions. \emph{Right}: The weighted average physics asymmetries for the two beam energies 4.725 GeV (blue) and  5.895 GeV (red) are shown. The dashed lines are the expected values of the physics asymmetries for the two beam energies calculated from the known form factor ratio for each Q$^2$ bin by Kelly's form factor parametrization \cite{99}.} 
\label{physasym}                  
\end{figure}

Figure~\ref{physasym} (\emph{right}) shows the extracted weighted average physics asymmetries for both beam energies for the coincidence data. The results are shown in Tab.~\ref{tabasym} (\emph{right half}). 

\subsection{Extraction of the \boldmath $G^p_E/G^p_M$ Ratio}
 
One can extract $\mu_p G_E^p/G_M^p$ for a known target spin orientation from the beam-target asymmetry in Eq.~(\ref{asym}) by solving for $R$. 

The four-momentum transfer squared, $Q^2(E, E', \theta_e)$, can be obtained for elastic events by knowing $\theta_e$ or $E'$ alone with equal accuracy from either quantity. However, propagating systematic uncertainties for $\theta_e (\delta \theta_e$=0.5 mrad$)$ and $E' (\delta E'/E'=0.1 \%)$ allows to evaluate the accuracy for determining $Q^2$ from $\theta_e$ or $E'$, respectively, and we found that it is more accurate to determine $Q^2$ from $\theta_e$.
%Since $Q^2$ from $\theta_e$is more accurate than $Q^2$ from $E'$, $Q^2$
Therefore, we used the electron angle, $\theta_e$, to calculate $Q^2$ for the  selected elastic events and found good agreement in the shape with the $Q^2$ distribution from the Monte Carlo simulation.

The mean values of the $Q^2$ distributions of events that pass the elastic cut on $W$ and the $\delta$ cuts are shown in Tab. \ref{tabasym}, and were used to calculate $\tau$, which appears in the terms $a,b,c$ in Eq.~(\ref{asym}).  
%The mean value of the $Q^2$ distribution was used to calculate $\tau$ which is used in the terms $a,b,c$ in Eq.~(\ref{asym}).  
The mean of the detected (or calculated using elastic kinematics of the proton in HMS) electron scattering angle, $\theta_e$ was determined by the $\theta_e$ distribution for the selected electrons of the single-arm (coincidence) data. The polarization angles, $\theta^*$ and $\phi ^*$, were calculated as
 \begin{eqnarray}
% \begin{aligned}
\label{thetastar}
\theta^*&=&\arccos (-\sin \theta_q\cos\phi_e\sin\beta+\cos\theta_q\cos
\beta) \\ 
%\end{equation}
%\begin{equation}
%\label{phistar}
\phi^*&=&-\arctan\left(\frac{\sin \phi_e \sin \beta}{\cos \theta_q
    \cos \phi_e \sin \beta + \sin \theta_q \cos
    \beta}\right)+180^{\circ} \nonumber. 
%\end{aligned}
\end{eqnarray}

The out-of-plane angle of the scattered electron at the target plane, $\phi_e$, is the mean of the detected $\phi_e$ distribution for the elastic events. The three-momentum transfer vector, $\vec{q}$, points at an angle $\theta_q$, which is identical with the elastically scattered proton angle, and is measured event-by-event for the elastic kinematics of the electron (proton) in the HMS. The mean value of the $\theta_q$ distribution was used in Eq.~(\ref{thetastar}). The target magnetic field direction was oriented with $\beta$=$80^{\circ}$ toward the BETA detector package from the beam line direction within the horizontal plane. The distribution of $\phi^*$ arises from the $\phi_e$ acceptance distribution. If $\phi_e=0$, then $\phi^*=0$ for single-arm data and $\phi^* = 180^{\circ}$ for coincidence data.  

The physics asymmetries, $A_p$, and the extracted proton form factor ratios, $R=G_E^p/G_M^p$, together with the average kinematic parameters for both single-arm and coincidence data are shown in Tab.~\ref{tabasym}. 
   \begin{table*}
%   \centering
   \begin{tabular}{|l|c|c|c|c|} 
      \hline 
         & \multicolumn{2}{c|}{single-arm} & \multicolumn{2}{c|}{Coincidence} \\
      \hline
         & $-8\% \textless \delta \textless 10\%$ & $10\% \textless \delta \textless 12\%$ & \multicolumn{2}{c|}{} \\
      \hline
       $E$            (GeV)                   & 5.895        & 5.895       &  5.893     &     4.725   \\
      $\theta_q$ (deg)            &   44.38      &   46.50     &   22.23     &    22.60   \\
      $\phi_q$    (deg)              & 171.80      & 172.20     & 188.40    &  190.90   \\      
      $\theta_e$ (deg)            &   15.45      &   14.92     &    37.08    &     43.52  \\      
      $\phi_e$    (deg)               & 351.80      & 352.10     &       8.40   &     10.95  \\
      $Q^2$ (GeV/c)$^2$           &      2.20      &      1.91    &       6.19   &        5.14  \\
      $\theta^*$ (deg)            &     36.31    &    34.20    &  101.90   &     102.10 \\
      $\phi^*$ (deg)                &   193.72    &  193.94    &       8.40   &       11.01 \\
      $A_p\pm \delta A_p$         & $-0.205 \pm 0.018$ & $-0.139 \pm 0.026$ & $0.083 \pm 0.074$ & $0.248 \pm 0.138$\\
      $\mu_p R \pm \delta (\mu_p R)$  & $0.576 \pm 0.217$  & $0.973 \pm 0.298$  & $0.439 \pm 0.411$   &  $-0.379\pm 0.690$\\
      \hline
      $A_p$ (expected)                 & $-0.186$ & $-0.171$  & $0.107$   &  $0.097$\\
      $\mu_p R$ (expected)            & $0.73$     & $0.78$      & $0.305$   &  $0.38$ \\
      \hline    
      \end{tabular}
   \caption{The experimental parameters together with the physics asymmetries and the extracted form factor ratios $\mu_p R = \mu_p G_E^p/G_M^p$ for both single-arm and coincidence data. The expected ratio $\mu_p R$ from Kelly's form factor parametrization \cite{99} for each $Q^2$ and the calculated asymmetry $A_p$ from the expected $\mu_p R$ are also shown. The errors $\delta A_p$ and $\delta(\mu_p R)$ are statistical.} 
   \label{tabasym}
   \end{table*}

\subsection{Systematic Error Estimation}

The systematic error of the form factor ratio, $\Delta (G_E^p/G_M^p)$, was determined by propagating the errors from the
experimental parameters to the physics asymmetry, $\Delta A_p$.  

The errors arising from the kinematic quantities were estimated by varying each quantity, one at a time by its corresponding uncertainty ($\delta E/E = 0.05$\% for the beam energy, $\delta P/P = 0.1$\% for the central momenta, and $\delta \theta_{e} = 0.5$ mrad for the spectrometer angle), and by propagating these errors to the $G_E^p/G_M^p$ ratios, which are extracted with the aid of the MC simulation. The resulting difference between the extracted $G_E^p/G_M^p$ ratio from the value at the nominal kinematics and the value shifted by the kinematic uncertainty was taken as the contribution to the systematic uncertainty in the $G_E^p/G_M^p$ ratio due to that quantity. In general, the uncertainties due to the kinematic variables, $E, E'(= P)$ and $\theta_e$ are less than 1\%.  

Using the Jacobian of the elastic electron-proton reaction, the error on the momentum transfer angle, $\delta \theta_q$, was obtained from $\delta E$ and the $\delta \theta_e$ and estimated as $\delta\theta_q = 0.03^{\circ}$. In addition, by assuming an error of the target magnetic field direction of $\delta \beta=0.1^{\circ}$, the uncertainties of $\theta^*$ and $\phi ^*$ were estimated to be $\delta \theta^* = 1.22$ mrad and $\delta \phi^* = 0.3$ mrad. The error of
$G_E^p/G_M^p$ from $\delta \theta^*$ was determined as 0.54\%, while that from $\delta \phi ^*$ was determined as 0.01\%. The systematic error on the target polarization was estimated as 5\%, which constitutes the largest systematic uncertainty \cite{109}. The error on the beam polarization measurement comes from a global error of the M$\o$ller measurements and the error due to the fit to these measurements. The beam polarization uncertainty during SANE was measured as 1.5\% \cite{109}. 

For both single-arm and coincidence data sets, the dilution factors have been determined using the comparison of data-to-Monte Carlo simulated yields. Since the simulated yields were based on the packing fraction, the error of 5\% on the packing fraction measurement propagates to the dilution factor. Therefore, the uncertainty of the form factor ratio, $G_E^p/G_M^p$, due to the error of the dilution factor was determined as 1.34\%. 

Single-arm data were analyzed using an extended momentum acceptance for the region of  $10\% \textless \delta \textless 12\%$, where the HMS optics were not well-tested. The reconstruction of the particle tracks from this region was not well-understood. Therefore, the uncertainty of the spectrometer optics on this region was a particular source of systematic uncertainty for the single-arm data \cite{147}. This has been tested with the Monte Carlo simulation. The
biggest loss of events in this higher $\delta$ region, $10\% \textless \delta \textless 12\%$, was found to be at the HMS vacuum pipe exit. By applying $\pm$2 mm offsets to the vacuum pipe positions on both vertical and horizontal directions separately in the MC simulation, and taking the standard effective solid angle change between the offset and the nominal vacuum pipe position, the uncertainty due to higher-momentum electron tracks hitting the edge of the vacuum pipe exit was determined.  The resulting uncertainty due to the particle track reconstruction and effective solid angle change was estimated as
0.68\%.  

Table~\ref{tabsys} summarizes non-negligible contributions to the systematic uncertainty of the single-arm data. Each source of systematics, the uncertainty of each quantity, and the resulting contribution to the relative systematic uncertainty of the  $\mu_p G_E^p/G_M^p$ ratio (=$\mu_p R$) are shown. The total uncorrelated relative systematic uncertainty was obtained by summing all the individual contributions quadratically and the final error on the form factor ratio was estimated as  5.44$\%$. The polarizations of the beam and target and the packing fraction were the dominant contributions to the systematic uncertainty. %However, due to lack of statistics for the
For the coincidence data, which are statistically limited, the systematic uncertainty was estimated based on the detailed systematics study at the single-arm data and found
%negligible. 
to be {$<0.1\%$}.

  \begin{table}[htbp]
   \centering
   \begin{tabular}{|l|c|c|} 
      \hline
         Quantity         & Error    & $\frac{\delta (\mu_p G_E^p/G_M^p)}{\mu_p G_E^p/G_M^p}$  \\
      \hline
      $E$ (GeV)           & 0.003    &   0.07\%   \\
      $E'$ (GeV)          & 0.004    &   0.13\%   \\      
      $\theta_e$ (mrad)   & 0.5      &   0.54\%   \\          
      $\theta^*$ (mrad)   & 1.22     &   0.54\%          \\
      $\phi^*$ (mrad)     & 0.3      &   0.01\%         \\      
      $P_T$   ($\%$)      & 5.0      &   5.0\%    \\
      $P_B$     ($\%$)    & 1.5      &   1.5\%     \\
      Packing Fraction, \emph{pf}   ($\%$)   &  5 &   1.34\%   \\
%      $10 \textless \delta \textless 12$   & 0.68 $\%$  &   \\            
 %     \hline
   %   \multicolumn{2}{|l|}{Linear sum :}     &   9.13\%    \\
      \hline
      \multicolumn{2}{|l|}{Quadratic sum : } &   5.44\%    \\
      \hline      
  \end{tabular}
   \caption{ Systematic uncertainty of each parameter and the relative systematic uncertainty on the $\mu_p G_E^p/G_M^p$ ratio due to the propagated uncertainty for the single-arm data. The maximum possible systematic uncertainty is obtained by the linear sum of all individual contributions. The final systematic uncertainty is obtained by the quadratic sum of all individual contributions.} 
   \label{tabsys}
   \end{table}

\section{Results}

The results for the proton elastic form factor ratio, $\mu_p G_E^p/G_M^p$, determined for both single-arm and coincidence data, are shown in Tab.~\ref{tabasym}. For the single-arm data, the resulting form factor ratio from the two $\delta$ regions of the HMS momentum acceptance was determined by extrapolating the short interval in $Q^2$ from the location of each of the two data points to the nominal location of the average of both. For the shape of the $Q^2$ dependence (or $Q^2$ evolution), the Kelly parametrization \cite{99} was used. After extrapolating each data point to the nominal average $Q^2$ location, the weighted average of both data points was taken. The resulting form factor ratio, $\mu_p G_E^p/G_M^p = 0.720 \pm 0.176_{stat} \pm 0.039_{sys}$ was obtained for an average four-momentum transfer squared $Q^2=2.06$ (GeV/c)$^2$. 

The form factor ratios from the coincidence data from two beam energies were also combined and the weighted average $\mu_p G_E^p/G_M^p$ was obtained at the average $Q^2=5.66$ (GeV/c)$^2$. Since the errors on the coincidence data were largely dominated by statistics, the systematic uncertainties  were not explicitly studied. Instead, the systematics from single-arm data were applied for an estimation. The resulting form factor ratio for the coincidence data was obtained as $\mu_p G_E^p/G_M^p = 0.244 \pm 0.353_{stat} \pm 0.013_{sys}$ for an average $Q^2=5.66$ (GeV/c)$^2$. 

Table~\ref{final} shows the final values for the $\mu_p G_E^p/G_M^p$ ratio together with the statistical and systematic uncertainties at each average
$Q^2$ value.   
  \begin{table}[htbp]
   \centering
   \begin{tabular}{|c|c|} 
      \hline
      $<Q^2>$ / (GeV/c)$^2$  & $\mu_p R \pm \delta (\mu_p R_{stat}) \pm \delta (\mu_p R_{sys})$\\
      \hline
      $2.06$                 &  0.720 $\pm$ 0.176 $\pm$ 0.039  \\ %adding the statistical and systematic errors in quadrature = 0.180
      $5.66$                 &  0.244 $\pm$ 0.353 $\pm$ 0.013  \\  %adding the statistical and systematic errors in quadrature  = 0.353 
      \hline
  \end{tabular}
     \vspace{-0.1cm}
   \caption{Results of the form factor analysis from the experiment SANE. The systematic error is based on the quadratic sum of individual contributions in Tab.~\ref{tabsys}.}
   \label{final}
   \end{table}

\begin{figure}[htbp]
\centering
\mbox{\includegraphics[width=\linewidth]{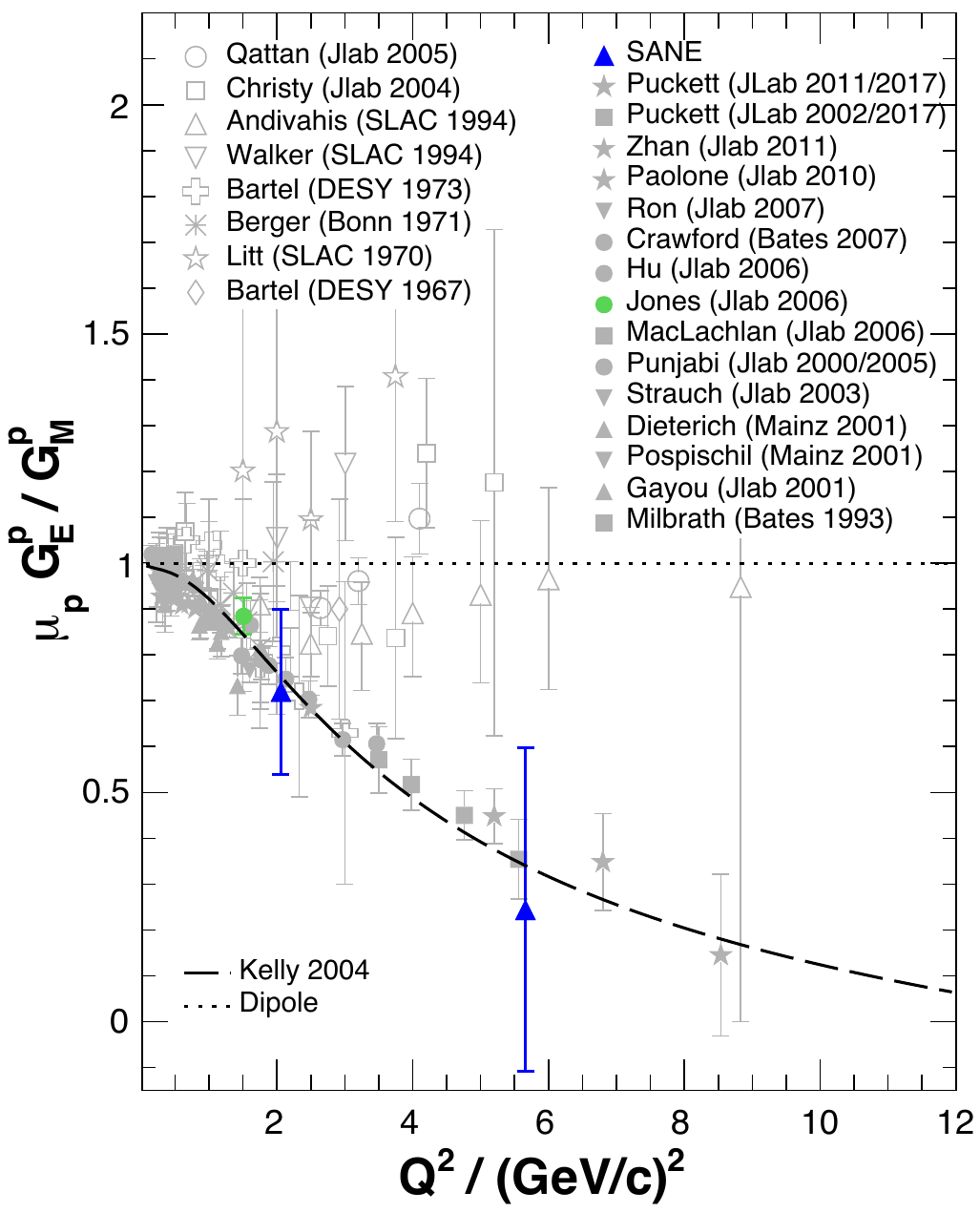}}
\caption{The form factor measurements from SANE together with the world data as a function of $Q^2$. Since the systematic errors are very small, only the total error bars, which are obtained by adding the statistical and systematic errors in quadrature, are shown. The previous polarized target experiment data is also highlighted (green) \cite{59}.} % and the systematic uncertainty
    
\label{resultratio}
\end{figure}

Figure~\ref{resultratio} shows the form factor measurements from SANE together with the world data as a function of $Q^2$. Since the systematic errors are very small, only the total error bars, which obtained by adding the statistical and systematic errors in quadrature are shown.

\section{Discussion and Conclusion}

Measurements of the proton's elastic form factor ratio, $\mu{_p}
G{^p}_E/G{^p}_M$, from the polarization-transfer experiments at high
$Q^2$ continue to show a dramatic discrepancy with the ratio obtained 
from the traditional Rosenbluth technique in unpolarized cross section
measurements as shown in Fig~\ref{resultratio}.  
The measurement of the beam-target asymmetry in the elastic \emph{ep}
scattering is an independent, third technique to determine the proton 
form factor ratio. 
The results from this method are in full agreement with the
proton recoil polarization data, which validates the
polarization-transfer method and reaffirms the discrepancy between
Rosenbluth and polarization data with different systematics.  
Two-photon exchange (TPE) continues to be a possible explanation for the
form factor discrepancy at high $Q^2$. 
However, the discrepancy may or may not be due to TPE, and further TPE 
measurements at high $Q^2$ need to be made before a final conclusion on 
TPE can be achieved.

Since the sensitivity to the form factor ratio and TPE effect is the same, this method was expected to
show consistent results with the recoil polarization method. Having
different systematic errors from the Rosenbluth method and the
polarization-transfer technique, the measurement of $G_E/G_M$ with the
polarized target technique has the potential to uncover unknown or underestimated 
systematic errors in the previous measurement techniques. 
 
Our result for $\mu{_p} G{^p}_E/G{^p}_M$ at $Q^2$=2.06 (GeV/c)$^2$ is 
consistent with the previous measurement of the beam-target asymmetry at 
$Q^2$=1.5 (GeV/c)$^2$~\cite{60} and agrees very well with the existing 
recoil-polarization measurements.
Our measurement did not reveal any unknown systematic difference from
the polarization-transfer method. 

The result at $Q^2$=5.66 (GeV/c)$^2$ has a larger statistical
uncertainty due to the small number of events. As a byproduct
measurement of the SANE experiment, the form factor measurement with HMS was not under optimized conditions and hence the precision of the result is
limited by statistics.
Furthermore, a gas leak in HMS drift chamber during the coincidence data taking resulted
in only 40\% efficiency for elastic proton detection with the
HMS. In addition, due to a damage of the superconducting Helmholtz
coils that were used to polarize the NH$_3$ target \cite{SANENIM}, the production
data-taking time was reduced.  
Therefore, single-arm data were taken for only about $\sim$12 hours in
total, while coincidence data for elastic kinematics were taken for
only about one week for both beam energies 4.725 GeV and 5.895 GeV,
$\sim$40 hours and $\sim$155 hours, respectively. The target spin
orientation was not optimized for the measurement of $G_E/G_M$.  
Nevertheless, the obtained precision confirms the
suitability of using the beam-target asymmetry for determinations of
the $\mu_p G_E^p/G_M^p$ ratio at high $Q^2$.  
% can we add an estimate how high in Q2 we can go for 10% error and 1000h?

Under optimized conditions, it would have been possible to take at least 
four times the amount of data in the same time period, which would have decreased
the error bars on both measurements by at least a factor of two. It is
hence suitable to extend the polarized-target technique to higher $Q^2$ and
achieve high precision with a dedicated experiment under optimized
conditions. 

\section{Acknowledgements}

%This work has been support with grants from DOE (DE-SC0003884, DE-SC-0013941 and DE-FG02-96ER40950), NSF (PHY-0855473, PHY-1207672, HRD-1649909) and Natural Sciences and Engineering Research Council of Canada (NSERC).
This work has been supported with grants from DOE (DE-SC0003884, DE-SC-0013941 and DE-FG02-96ER40950), NSF (PHY-0855473, PHY-1207672, HRD-1649909) and Natural Sciences and Engineering Research Council of Canada (NSERC).  This material is based upon work supported by the U.S. Department of Energy, Office of Science, Office of Nuclear Physics under contract DE-AC05-06OR23177.

\bibliography{ref_new}

\begin{thebibliography}{73}
\expandafter\ifx\csname natexlab\endcsname\relax\def\natexlab#1{#1}\fi
\expandafter\ifx\csname bibnamefont\endcsname\relax
  \def\bibnamefont#1{#1}\fi
\expandafter\ifx\csname bibfnamefont\endcsname\relax
  \def\bibfnamefont#1{#1}\fi
\expandafter\ifx\csname citenamefont\endcsname\relax
  \def\citenamefont#1{#1}\fi
\expandafter\ifx\csname url\endcsname\relax
  \def\url#1{\texttt{#1}}\fi
\expandafter\ifx\csname urlprefix\endcsname\relax\def\urlprefix{URL }\fi
\providecommand{\bibinfo}[2]{#2}
\providecommand{\eprint}[2][]{\url{#2}}

\bibitem[{\citenamefont{Rosenbluth}(1950)}]{rosenbluth1950}
\bibinfo{author}{\bibfnamefont{M.~N.} \bibnamefont{Rosenbluth}},
  \bibinfo{journal}{Phys. Rev.} \textbf{\bibinfo{volume}{79}},
  \bibinfo{pages}{615} (\bibinfo{year}{1950}),
  \urlprefix\url{https://journals.aps.org/pr/abstract/10.1103/PhysRev.79.615}.

\bibitem[{\citenamefont{Akhiezer and Rekalo}(1968)}]{157}
\bibinfo{author}{\bibfnamefont{A.~I.} \bibnamefont{Akhiezer}} \bibnamefont{and}
  \bibinfo{author}{\bibfnamefont{M.~P.} \bibnamefont{Rekalo}},
  \bibinfo{journal}{Sov. Phys. Dokl.} \textbf{\bibinfo{volume}{13}},
  \bibinfo{pages}{572} (\bibinfo{year}{1968}), \bibinfo{note}{[Dokl. Akad. Nauk
  Ser. Fiz.180,1081(1968)]}.

\bibitem[{\citenamefont{Arnold et~al.}(1981)\citenamefont{Arnold, Carlson, and
  Gross}}]{160}
\bibinfo{author}{\bibfnamefont{R.~G.} \bibnamefont{Arnold}},
  \bibinfo{author}{\bibfnamefont{C.~E.} \bibnamefont{Carlson}},
  \bibnamefont{and} \bibinfo{author}{\bibfnamefont{F.}~\bibnamefont{Gross}},
  \bibinfo{journal}{Phys. Rev.} \textbf{\bibinfo{volume}{C23}},
  \bibinfo{pages}{363} (\bibinfo{year}{1981}),
  \urlprefix\url{https://journals.aps.org/prc/abstract/10.1103/PhysRevC.23.363}.

\bibitem[{\citenamefont{Dombey}(1969)}]{158}
\bibinfo{author}{\bibfnamefont{N.}~\bibnamefont{Dombey}},
  \bibinfo{journal}{Rev. Mod. Phys.} \textbf{\bibinfo{volume}{41}},
  \bibinfo{pages}{236} (\bibinfo{year}{1969}),
  \urlprefix\url{https://journals.aps.org/rmp/abstract/10.1103/RevModPhys.41.236}.

\bibitem[{\citenamefont{Donnelly and Raskin}(1986)}]{156}
\bibinfo{author}{\bibfnamefont{T.~W.} \bibnamefont{Donnelly}} \bibnamefont{and}
  \bibinfo{author}{\bibfnamefont{A.~S.} \bibnamefont{Raskin}},
  \bibinfo{journal}{Annals Phys.} \textbf{\bibinfo{volume}{169}},
  \bibinfo{pages}{247} (\bibinfo{year}{1986}).

\bibitem[{\citenamefont{{I. A. Qattan \emph{et al.}}}(2005)}]{26}
\bibinfo{author}{\bibnamefont{{I. A. Qattan \emph{et al.}}}},
  \bibinfo{journal}{Phys. Rev. Lett.} \textbf{\bibinfo{volume}{94}},
  \bibinfo{pages}{142301} (\bibinfo{year}{2005}), \eprint{nucl-ex/0410010},
  \urlprefix\url{https://journals.aps.org/prl/abstract/10.1103/PhysRevLett.94.142301}.

\bibitem[{\citenamefont{{M. E. Christy \emph{et al.}}}(2004)}]{27}
\bibinfo{author}{\bibnamefont{{M. E. Christy \emph{et al.}}}}
  (\bibinfo{collaboration}{E94110}), \bibinfo{journal}{Phys. Rev.}
  \textbf{\bibinfo{volume}{C70}}, \bibinfo{pages}{015206}
  (\bibinfo{year}{2004}), \eprint{nucl-ex/0401030},
  \urlprefix\url{https://journals.aps.org/prc/abstract/10.1103/PhysRevC.70.015206}.

\bibitem[{\citenamefont{{L. Andivahis \emph{et al.}}}(1994)}]{28}
\bibinfo{author}{\bibnamefont{{L. Andivahis \emph{et al.}}}},
  \bibinfo{journal}{Phys. Rev.} \textbf{\bibinfo{volume}{D50}},
  \bibinfo{pages}{5491} (\bibinfo{year}{1994}),
  \urlprefix\url{https://journals.aps.org/prd/abstract/10.1103/PhysRevD.50.5491}.

\bibitem[{\citenamefont{{R. C. Walker \emph{et al.}}}(1994)}]{29}
\bibinfo{author}{\bibnamefont{{R. C. Walker \emph{et al.}}}},
  \bibinfo{journal}{Phys. Rev.} \textbf{\bibinfo{volume}{D49}},
  \bibinfo{pages}{5671} (\bibinfo{year}{1994}),
  \urlprefix\url{https://journals.aps.org/prd/abstract/10.1103/PhysRevD.49.5671}.

\bibitem[{\citenamefont{{F. Borkowski \emph{et al.}}}(1975)}]{30}
\bibinfo{author}{\bibnamefont{{F. Borkowski \emph{et al.}}}},
  \bibinfo{journal}{Nucl. Phys.} \textbf{\bibinfo{volume}{B93}},
  \bibinfo{pages}{461} (\bibinfo{year}{1975}).

\bibitem[{\citenamefont{{F. Borkowski \emph{et al.}}}(1974)}]{31}
\bibinfo{author}{\bibnamefont{{F. Borkowski \emph{et al.}}}},
  \bibinfo{journal}{Nucl. Phys.} \textbf{\bibinfo{volume}{A222}},
  \bibinfo{pages}{269} (\bibinfo{year}{1974}).

\bibitem[{\citenamefont{{W. Bartel \emph{et al.}}}(1973)}]{32}
\bibinfo{author}{\bibnamefont{{W. Bartel \emph{et al.}}}},
  \bibinfo{journal}{Nucl. Phys.} \textbf{\bibinfo{volume}{B58}},
  \bibinfo{pages}{429} (\bibinfo{year}{1973}).

\bibitem[{\citenamefont{{C. Berger \emph{et al.}}}(1971)}]{33}
\bibinfo{author}{\bibnamefont{{C. Berger \emph{et al.}}}},
  \bibinfo{journal}{Phys. Lett.} \textbf{\bibinfo{volume}{35B}},
  \bibinfo{pages}{87} (\bibinfo{year}{1971}).

\bibitem[{\citenamefont{{J. Litt \emph{et. al.}}}(1970)}]{34}
\bibinfo{author}{\bibnamefont{{J. Litt \emph{et. al.}}}},
  \bibinfo{journal}{Phys. Lett.} \textbf{\bibinfo{volume}{31B}},
  \bibinfo{pages}{40} (\bibinfo{year}{1970}).

\bibitem[{\citenamefont{{T. Janssens \emph{et al.}}}(1966)}]{35}
\bibinfo{author}{\bibnamefont{{T. Janssens \emph{et al.}}}},
  \bibinfo{journal}{Phys. Rev.} \textbf{\bibinfo{volume}{142}},
  \bibinfo{pages}{922} (\bibinfo{year}{1966}),
  \urlprefix\url{https://journals.aps.org/pr/abstract/10.1103/PhysRev.142.922}.

\bibitem[{\citenamefont{{M. Meziane \emph{et al.}}}(2011 (superseded by
  \cite{Puckett:2017flj}))}]{43}
\bibinfo{author}{\bibnamefont{{M. Meziane \emph{et al.}}}}
  (\bibinfo{collaboration}{GEp2$\gamma$ Collaboration}),
  \bibinfo{journal}{Phys. Rev. Lett.} \textbf{\bibinfo{volume}{106}},
  \bibinfo{pages}{132501} (\bibinfo{year}{2011 (superseded by
  \cite{Puckett:2017flj})}),
  \urlprefix\url{https://journals.aps.org/prl/abstract/10.1103/PhysRevLett.106.132501}.

\bibitem[{\citenamefont{{A. J. R. Puckett \emph{et al.}}}(2010 (superseded by
  \cite{Puckett:2017flj}))}]{44}
\bibinfo{author}{\bibnamefont{{A. J. R. Puckett \emph{et al.}}}},
  \bibinfo{journal}{Phys. Rev. Lett.} \textbf{\bibinfo{volume}{104}},
  \bibinfo{pages}{242301} (\bibinfo{year}{2010 (superseded by
  \cite{Puckett:2017flj})}),
  \urlprefix\url{http://link.aps.org/doi/10.1103/PhysRevLett.104.242301}.

\bibitem[{\citenamefont{{M. Paolone \emph{et al.}}}(2010)}]{45}
\bibinfo{author}{\bibnamefont{{M. Paolone \emph{et al.}}}},
  \bibinfo{journal}{Phys. Rev. Lett.} \textbf{\bibinfo{volume}{105}},
  \bibinfo{pages}{072001} (\bibinfo{year}{2010}), \eprint{1002.2188},
  \urlprefix\url{https://journals.aps.org/prl/abstract/10.1103/PhysRevLett.105.072001}.

\bibitem[{\citenamefont{{G. Ron \emph{et al.}}}(2007)}]{46}
\bibinfo{author}{\bibnamefont{{G. Ron \emph{et al.}}}}, \bibinfo{journal}{Phys.
  Rev. Lett.} \textbf{\bibinfo{volume}{99}}, \bibinfo{pages}{202002}
  (\bibinfo{year}{2007}), \eprint{0706.0128},
  \urlprefix\url{https://journals.aps.org/prl/abstract/10.1103/PhysRevLett.99.202002}.

\bibitem[{\citenamefont{{B. Hu and M. K. Jones \emph{et al.}}}(2006)}]{47}
\bibinfo{author}{\bibnamefont{{B. Hu and M. K. Jones \emph{et al.}}}},
  \bibinfo{journal}{Phys. Rev.} \textbf{\bibinfo{volume}{C73}},
  \bibinfo{pages}{064004} (\bibinfo{year}{2006}), \eprint{nucl-ex/0601025},
  \urlprefix\url{https://journals.aps.org/prc/abstract/10.1103/PhysRevC.73.064004}.

\bibitem[{\citenamefont{{G. MacLachlan \emph{et al.}}}(2006)}]{48}
\bibinfo{author}{\bibnamefont{{G. MacLachlan \emph{et al.}}}},
  \bibinfo{journal}{Nucl. Phys.} \textbf{\bibinfo{volume}{A764}},
  \bibinfo{pages}{261} (\bibinfo{year}{2006}).

\bibitem[{\citenamefont{{V. Punjabi and C. F. Perdrisat \emph{et al.}
  }}(2005)}]{50}
\bibinfo{author}{\bibnamefont{{V. Punjabi and C. F. Perdrisat \emph{et al.} }}}
  (\bibinfo{collaboration}{Jefferson Lab Hall A Collaboration}),
  \bibinfo{journal}{Phys. Rev.} \textbf{\bibinfo{volume}{C71}},
  \bibinfo{pages}{055202} (\bibinfo{year}{2005}), \bibinfo{note}{[Erratum:
  Phys. Rev. {\bf C71}, 069902 (2005)]},
  \urlprefix\url{https://journals.aps.org/prc/abstract/10.1103/PhysRevC.71.055202}.

\bibitem[{\citenamefont{{S. Strauch \emph{et al.}}}(2003)}]{51}
\bibinfo{author}{\bibnamefont{{S. Strauch \emph{et al.}}}}
  (\bibinfo{collaboration}{Jefferson Lab E93-049}), \bibinfo{journal}{Phys.
  Rev. Lett.} \textbf{\bibinfo{volume}{91}}, \bibinfo{pages}{052301}
  (\bibinfo{year}{2003}), \eprint{nucl-ex/0211022},
  \urlprefix\url{https://journals.aps.org/prl/abstract/10.1103/PhysRevLett.91.052301}.

\bibitem[{\citenamefont{{O. Gayou \emph{et al.}}}(2002 (superseded by
  \cite{Puckett:2011xg}))}]{52}
\bibinfo{author}{\bibnamefont{{O. Gayou \emph{et al.}}}}
  (\bibinfo{collaboration}{Jefferson Lab Hall A}), \bibinfo{journal}{Phys. Rev.
  Lett.} \textbf{\bibinfo{volume}{88}}, \bibinfo{pages}{092301}
  (\bibinfo{year}{2002 (superseded by \cite{Puckett:2011xg})}),
  \eprint{nucl-ex/0111010},
  \urlprefix\url{https://journals.aps.org/prl/abstract/10.1103/PhysRevLett.88.092301}.

\bibitem[{\citenamefont{{O. Gayou \emph{et al.}}}(2001)}]{53}
\bibinfo{author}{\bibnamefont{{O. Gayou \emph{et al.}}}},
  \bibinfo{journal}{Phys. Rev.} \textbf{\bibinfo{volume}{C64}},
  \bibinfo{pages}{038202} (\bibinfo{year}{2001}),
  \urlprefix\url{https://journals.aps.org/prc/abstract/10.1103/PhysRevC.64.038202}.

\bibitem[{\citenamefont{{S. Dieterich \emph{et al.}}}(2001)}]{54}
\bibinfo{author}{\bibnamefont{{S. Dieterich \emph{et al.}}}},
  \bibinfo{journal}{Phys. Lett.} \textbf{\bibinfo{volume}{B500}},
  \bibinfo{pages}{47} (\bibinfo{year}{2001}), \eprint{nucl-ex/0011008}.

\bibitem[{\citenamefont{{T. Pospischil \emph{et al.}}}(2001)}]{55}
\bibinfo{author}{\bibnamefont{{T. Pospischil \emph{et al.}}}}
  (\bibinfo{collaboration}{A1 Collaboration}), \bibinfo{journal}{Eur. Phys. J.}
  \textbf{\bibinfo{volume}{A12}}, \bibinfo{pages}{125} (\bibinfo{year}{2001}).

\bibitem[{\citenamefont{{M. K. Jones \emph{et al.}}}(2000 (superseded by
  \cite{50}))}]{56}
\bibinfo{author}{\bibnamefont{{M. K. Jones \emph{et al.}}}}
  (\bibinfo{collaboration}{Jefferson Lab Hall A Collaboration}),
  \bibinfo{journal}{Phys. Rev. Lett.} \textbf{\bibinfo{volume}{84}},
  \bibinfo{pages}{1398} (\bibinfo{year}{2000 (superseded by \cite{50})}),
  \eprint{nucl-ex/9910005},
  \urlprefix\url{https://journals.aps.org/prl/abstract/10.1103/PhysRevLett.84.1398}.

\bibitem[{\citenamefont{{B. D. Milbrath \emph{et al.}}}(1998)}]{58}
\bibinfo{author}{\bibnamefont{{B. D. Milbrath \emph{et al.}}}}
  (\bibinfo{collaboration}{Bates FPP collaboration}), \bibinfo{journal}{Phys.
  Rev. Lett.} \textbf{\bibinfo{volume}{80}}, \bibinfo{pages}{452}
  (\bibinfo{year}{1998}), \bibinfo{note}{[Erratum: Phys. Rev. Lett. {\bf 82},
  2221 (1999)]}, \eprint{nucl-ex/9712006},
  \urlprefix\url{https://journals.aps.org/prl/abstract/10.1103/PhysRevLett.80.452}.

\bibitem[{\citenamefont{Puckett et~al.}(2017)}]{Puckett:2017flj}
\bibinfo{author}{\bibfnamefont{A.~J.~R.} \bibnamefont{Puckett}}
  \bibnamefont{et~al.}, \bibinfo{journal}{Phys. Rev.}
  \textbf{\bibinfo{volume}{C96}}, \bibinfo{pages}{055203}
  (\bibinfo{year}{2017}), \eprint{1707.08587},
  \urlprefix\url{https://journals.aps.org/prc/abstract/10.1103/PhysRevC.96.055203}.

\bibitem[{\citenamefont{Puckett et~al.}(2012)}]{Puckett:2011xg}
\bibinfo{author}{\bibfnamefont{A.~J.~R.} \bibnamefont{Puckett}}
  \bibnamefont{et~al.}, \bibinfo{journal}{Phys. Rev.}
  \textbf{\bibinfo{volume}{C85}}, \bibinfo{pages}{045203}
  (\bibinfo{year}{2012}), \eprint{1102.5737},
  \urlprefix\url{https://journals.aps.org/prc/abstract/10.1103/PhysRevC.85.045203}.

\bibitem[{\citenamefont{{M. K. Jones \emph{et al.}}}(2006)}]{59}
\bibinfo{author}{\bibnamefont{{M. K. Jones \emph{et al.}}}}
  (\bibinfo{collaboration}{Resonance Spin Structure Collaboration}),
  \bibinfo{journal}{Phys. Rev.} \textbf{\bibinfo{volume}{C74}},
  \bibinfo{pages}{035201} (\bibinfo{year}{2006}),
  \urlprefix\url{https://link.aps.org/doi/10.1103/PhysRevC.74.035201}.

\bibitem[{\citenamefont{{C. B. Crawford \emph{et al.}}}(2007)}]{60}
\bibinfo{author}{\bibnamefont{{C. B. Crawford \emph{et al.}}}},
  \bibinfo{journal}{Phys. Rev. Lett.} \textbf{\bibinfo{volume}{98}},
  \bibinfo{pages}{052301} (\bibinfo{year}{2007}),
  \urlprefix\url{http://link.aps.org/doi/10.1103/PhysRevLett.98.052301}.

\bibitem[{\citenamefont{Kelly}(2004)}]{99}
\bibinfo{author}{\bibfnamefont{J.~J.} \bibnamefont{Kelly}},
  \bibinfo{journal}{Phys. Rev.} \textbf{\bibinfo{volume}{C70}},
  \bibinfo{pages}{068202} (\bibinfo{year}{2004}),
  \urlprefix\url{https://journals.aps.org/prc/abstract/10.1103/PhysRevC.70.068202}.

\bibitem[{\citenamefont{Guichon and Vanderhaeghen}(2003)}]{97}
\bibinfo{author}{\bibfnamefont{P.~A.~M.} \bibnamefont{Guichon}}
  \bibnamefont{and}
  \bibinfo{author}{\bibfnamefont{M.}~\bibnamefont{Vanderhaeghen}},
  \bibinfo{journal}{Phys. Rev. Lett.} \textbf{\bibinfo{volume}{91}},
  \bibinfo{pages}{142303} (\bibinfo{year}{2003}), \eprint{hep-ph/0306007},
  \urlprefix\url{https://journals.aps.org/prl/abstract/10.1103/PhysRevLett.91.142303}.

\bibitem[{\citenamefont{Blunden et~al.}(2003)\citenamefont{Blunden,
  Melnitchouk, and Tjon}}]{98}
\bibinfo{author}{\bibfnamefont{P.~G.} \bibnamefont{Blunden}},
  \bibinfo{author}{\bibfnamefont{W.}~\bibnamefont{Melnitchouk}},
  \bibnamefont{and} \bibinfo{author}{\bibfnamefont{J.~A.} \bibnamefont{Tjon}},
  \bibinfo{journal}{Phys. Rev. Lett.} \textbf{\bibinfo{volume}{91}},
  \bibinfo{pages}{142304} (\bibinfo{year}{2003}), \eprint{nucl-th/0306076},
  \urlprefix\url{https://journals.aps.org/prl/abstract/10.1103/PhysRevLett.91.142304}.

\bibitem[{\citenamefont{Rekalo and Tomasi-Gustafsson}(2004)}]{96}
\bibinfo{author}{\bibfnamefont{M.~P.} \bibnamefont{Rekalo}} \bibnamefont{and}
  \bibinfo{author}{\bibfnamefont{E.}~\bibnamefont{Tomasi-Gustafsson}},
  \bibinfo{journal}{Eur. Phys. J.} \textbf{\bibinfo{volume}{A22}},
  \bibinfo{pages}{331} (\bibinfo{year}{2004}), \eprint{nucl-th/0307066}.

\bibitem[{\citenamefont{Chen et~al.}(2004)\citenamefont{Chen, Afanasev,
  Brodsky, Carlson, and Vanderhaeghen}}]{95}
\bibinfo{author}{\bibfnamefont{Y.~C.} \bibnamefont{Chen}},
  \bibinfo{author}{\bibfnamefont{A.}~\bibnamefont{Afanasev}},
  \bibinfo{author}{\bibfnamefont{S.~J.} \bibnamefont{Brodsky}},
  \bibinfo{author}{\bibfnamefont{C.~E.} \bibnamefont{Carlson}},
  \bibnamefont{and}
  \bibinfo{author}{\bibfnamefont{M.}~\bibnamefont{Vanderhaeghen}},
  \bibinfo{journal}{Phys. Rev. Lett.} \textbf{\bibinfo{volume}{93}},
  \bibinfo{pages}{122301} (\bibinfo{year}{2004}), \eprint{hep-ph/0403058},
  \urlprefix\url{https://arxiv.org/pdf/hep-ph/0403058.pdf}.

\bibitem[{\citenamefont{Afanasev et~al.}(2005)\citenamefont{Afanasev, Brodsky,
  Carlson, Chen, and Vanderhaeghen}}]{94}
\bibinfo{author}{\bibfnamefont{A.~V.} \bibnamefont{Afanasev}},
  \bibinfo{author}{\bibfnamefont{S.~J.} \bibnamefont{Brodsky}},
  \bibinfo{author}{\bibfnamefont{C.~E.} \bibnamefont{Carlson}},
  \bibinfo{author}{\bibfnamefont{Y.-C.} \bibnamefont{Chen}}, \bibnamefont{and}
  \bibinfo{author}{\bibfnamefont{M.}~\bibnamefont{Vanderhaeghen}},
  \bibinfo{journal}{Phys. Rev.} \textbf{\bibinfo{volume}{D72}},
  \bibinfo{pages}{013008} (\bibinfo{year}{2005}), \eprint{hep-ph/0502013},
  \urlprefix\url{https://journals.aps.org/prd/abstract/10.1103/PhysRevD.72.013008}.

\bibitem[{\citenamefont{Blunden et~al.}(2005)\citenamefont{Blunden,
  Melnitchouk, and Tjon}}]{93}
\bibinfo{author}{\bibfnamefont{P.~G.} \bibnamefont{Blunden}},
  \bibinfo{author}{\bibfnamefont{W.}~\bibnamefont{Melnitchouk}},
  \bibnamefont{and} \bibinfo{author}{\bibfnamefont{J.~A.} \bibnamefont{Tjon}},
  \bibinfo{journal}{Phys. Rev.} \textbf{\bibinfo{volume}{C72}},
  \bibinfo{pages}{034612} (\bibinfo{year}{2005}), \eprint{nucl-th/0506039},
  \urlprefix\url{https://journals.aps.org/prc/abstract/10.1103/PhysRevC.72.034612}.

\bibitem[{\citenamefont{{D. Borisyuk and A. Kobushkin}}(2006)}]{92}
\bibinfo{author}{\bibnamefont{{D. Borisyuk and A. Kobushkin}}},
  \bibinfo{journal}{Phys. Rev.} \textbf{\bibinfo{volume}{C74}},
  \bibinfo{pages}{065203} (\bibinfo{year}{2006}), \eprint{nucl-th/0606030},
  \urlprefix\url{https://journals.aps.org/prc/abstract/10.1103/PhysRevC.74.065203}.

\bibitem[{\citenamefont{{D. Borisyuk and A. Kobushkin}}(2008)}]{91}
\bibinfo{author}{\bibnamefont{{D. Borisyuk and A. Kobushkin}}},
  \bibinfo{journal}{Phys. Rev.} \textbf{\bibinfo{volume}{C78}},
  \bibinfo{pages}{025208} (\bibinfo{year}{2008}), \eprint{0804.4128},
  \urlprefix\url{https://journals.aps.org/prc/abstract/10.1103/PhysRevC.78.025208}.

\bibitem[{\citenamefont{{D. Borisyuk and A. Kobushkin}}(2009)}]{90}
\bibinfo{author}{\bibnamefont{{D. Borisyuk and A. Kobushkin}}},
  \bibinfo{journal}{Phys. Rev.} \textbf{\bibinfo{volume}{D79}},
  \bibinfo{pages}{034001} (\bibinfo{year}{2009}), \eprint{0811.0266},
  \urlprefix\url{https://journals.aps.org/prd/abstract/10.1103/PhysRevD.79.034001}.

\bibitem[{\citenamefont{{N. Kivel and M. Vanderhaeghen}}(2009)}]{89}
\bibinfo{author}{\bibnamefont{{N. Kivel and M. Vanderhaeghen}}},
  \bibinfo{journal}{Phys. Rev. Lett.} \textbf{\bibinfo{volume}{103}},
  \bibinfo{pages}{092004} (\bibinfo{year}{2009}), \eprint{0905.0282},
  \urlprefix\url{https://journals.aps.org/prl/abstract/10.1103/PhysRevLett.103.092004}.

\bibitem[{\citenamefont{Kondratyuk et~al.}(2005)\citenamefont{Kondratyuk,
  Blunden, Melnitchouk, and Tjon}}]{Kondratyuk:2005kk}
\bibinfo{author}{\bibfnamefont{S.}~\bibnamefont{Kondratyuk}},
  \bibinfo{author}{\bibfnamefont{P.~G.} \bibnamefont{Blunden}},
  \bibinfo{author}{\bibfnamefont{W.}~\bibnamefont{Melnitchouk}},
  \bibnamefont{and} \bibinfo{author}{\bibfnamefont{J.~A.} \bibnamefont{Tjon}},
  \bibinfo{journal}{Phys. Rev. Lett.} \textbf{\bibinfo{volume}{95}},
  \bibinfo{pages}{172503} (\bibinfo{year}{2005}), \eprint{nucl-th/0506026}.

\bibitem[{\citenamefont{{E. A. Kuraev and V. V. Bytev and S. Bakmaev and E.
  Tomasi-Gustafsson}}(2008)}]{Kuraev:2007dn}
\bibinfo{author}{\bibnamefont{{E. A. Kuraev and V. V. Bytev and S. Bakmaev and
  E. Tomasi-Gustafsson}}}, \bibinfo{journal}{Phys. Rev.}
  \textbf{\bibinfo{volume}{C78}}, \bibinfo{pages}{015205}
  (\bibinfo{year}{2008}), \eprint{0710.3699},
  \urlprefix\url{https://journals.aps.org/prc/abstract/10.1103/PhysRevC.78.015205}.

\bibitem[{\citenamefont{{S. Pacetti and E.
  Tomasi-Gustafsson}}(2016)}]{Pacetti:2016}
\bibinfo{author}{\bibnamefont{{S. Pacetti and E. Tomasi-Gustafsson}}},
  \bibinfo{journal}{Phys. Rev.} \textbf{\bibinfo{volume}{C94}},
  \bibinfo{pages}{055202} (\bibinfo{year}{2016}), \eprint{1604.02421},
  \urlprefix\url{https://journals.aps.org/prc/abstract/10.1103/PhysRevC.94.055202}.

\bibitem[{\citenamefont{{J. Arrington \emph{et al.}}}({2005})}]{163}
\bibinfo{author}{\bibnamefont{{J. Arrington \emph{et al.}}}},
  \bibinfo{journal}{Jefferson Lab Proposal E05-017}  (\bibinfo{year}{{2005}}),
  \urlprefix\url{https://www.jlab.org/exp_prog/CEBAF_EXP/E05017.html}.

\bibitem[{\citenamefont{{D. Adikaram \emph{et al.}}}(2017)}]{154}
\bibinfo{author}{\bibnamefont{{D. Adikaram \emph{et al.}}}}
  (\bibinfo{collaboration}{CLAS Collaboration}), \bibinfo{journal}{Phys. Rev.
  Lett.} \textbf{\bibinfo{volume}{114}}, \bibinfo{pages}{062003}
  (\bibinfo{year}{2017}).

\bibitem[{\citenamefont{{D. Rimal \emph{et al.}}}(2017)}]{161}
\bibinfo{author}{\bibnamefont{{D. Rimal \emph{et al.}}}}
  (\bibinfo{collaboration}{CLAS Collaboration}), \bibinfo{journal}{Phys. Rev.}
  \textbf{\bibinfo{volume}{C95}}, \bibinfo{pages}{065201}
  (\bibinfo{year}{2017}), \eprint{1603.00315},
  \urlprefix\url{https://journals.aps.org/prc/abstract/10.1103/PhysRevC.95.065201}.

\bibitem[{\citenamefont{{I. A. Rachek \emph{et al.}}}(2015)}]{162}
\bibinfo{author}{\bibnamefont{{I. A. Rachek \emph{et al.}}}},
  \bibinfo{journal}{Phys. Rev. Lett.} \textbf{\bibinfo{volume}{114}},
  \bibinfo{pages}{062005} (\bibinfo{year}{2015}), \eprint{1411.7372},
  \urlprefix\url{https://journals.aps.org/prl/abstract/10.1103/PhysRevLett.114.062005}.

\bibitem[{\citenamefont{{B. S. Henderson \emph{et al.}}}(2017)}]{155}
\bibinfo{author}{\bibnamefont{{B. S. Henderson \emph{et al.}}}}
  (\bibinfo{collaboration}{OLYMPUS Collaboration}), \bibinfo{journal}{Phys.
  Rev. Lett.} \textbf{\bibinfo{volume}{118}}, \bibinfo{pages}{092501}
  (\bibinfo{year}{2017}), \eprint{1611.04685},
  \urlprefix\url{https://journals.aps.org/prl/abstract/10.1103/PhysRevLett.118.092501}.

\bibitem[{\citenamefont{{J. Jourdan \emph{et al.}}}({2007})}]{164}
\bibinfo{author}{\bibnamefont{{J. Jourdan \emph{et al.}}}},
  \bibinfo{journal}{Jefferson Lab Proposal E07-003}  (\bibinfo{year}{{2007}}),
  \urlprefix\url{https://www.jlab.org/exp_prog/CEBAF_EXP/E07003.html}.

\bibitem[{\citenamefont{Maxwell}(2017)}]{109}
\bibinfo{author}{\bibfnamefont{J.~D.} \bibnamefont{Maxwell}},
  \bibinfo{journal}{Ph.D. thesis, University of Virginia, Charlottesville,
  Virginia}  (\bibinfo{year}{2017}), \eprint{1704.02308},
  \urlprefix\url{https://inspirehep.net/record/1590296/files/arXiv:1704.02308.pdf}.

\bibitem[{\citenamefont{Mulholland}(2012)}]{144}
\bibinfo{author}{\bibfnamefont{J.}~\bibnamefont{Mulholland}},
  \bibinfo{journal}{Ph.D. thesis, University Of Virginia, Charlottesville,
  Virginia}  (\bibinfo{year}{2012}),
  \urlprefix\url{https://misportal.jlab.org/ul/publications/view_pub.cfm?pub_id=11458}.

\bibitem[{\citenamefont{Liyanage}(2013)}]{142}
\bibinfo{author}{\bibfnamefont{A.~P.~H.} \bibnamefont{Liyanage}},
  \bibinfo{journal}{Ph.D. thesis, Hampton U.}  (\bibinfo{year}{2013}),
  \urlprefix\url{https://misportal.jlab.org/ul/publications/view_pub.cfm?pub_id=12790}.

\bibitem[{\citenamefont{Ndukum}(2015)}]{143}
\bibinfo{author}{\bibfnamefont{L.~Z.} \bibnamefont{Ndukum}},
  \bibinfo{journal}{Ph.D. thesis, Mississippi State U.}
  (\bibinfo{year}{2015}),
  \urlprefix\url{https://misportal.jlab.org/ul/publications/view_pub.cfm?pub_id=13854}.

\bibitem[{\citenamefont{Armstrong}(2015)}]{139}
\bibinfo{author}{\bibfnamefont{W.~R.} \bibnamefont{Armstrong}},
  \bibinfo{journal}{Ph.D. thesis, Temple University, Philadelphia}
  (\bibinfo{year}{2015}),
  \urlprefix\url{https://misportal.jlab.org/ul/publications/view_pub.cfm?pub_id=13921}.

\bibitem[{\citenamefont{Kang}(2015)}]{140}
\bibinfo{author}{\bibfnamefont{H.}~\bibnamefont{Kang}}, \bibinfo{journal}{Ph.D.
  thesis, Seoul National University, Seoul, South Korea}
  (\bibinfo{year}{2015}),
  \urlprefix\url{https://misportal.jlab.org/ul/publications/view_pub.cfm?pub_id=13922}.

\bibitem[{\citenamefont{{F. R. Wesselmann \emph{et al.}}}(2007)}]{179}
\bibinfo{author}{\bibnamefont{{F. R. Wesselmann \emph{et al.}}}}
  (\bibinfo{collaboration}{RSS}), \bibinfo{journal}{Phys. Rev. Lett.}
  \textbf{\bibinfo{volume}{98}}, \bibinfo{pages}{132003}
  (\bibinfo{year}{2007}), \eprint{nucl-ex/0608003},
  \urlprefix\url{https://journals.aps.org/prl/abstract/10.1103/PhysRevLett.98.132003}.

\bibitem[{\citenamefont{{O. Rondon \emph{et al.}}}({2017})}]{SANE2017}
\bibinfo{author}{\bibnamefont{{O. Rondon \emph{et al.}}}}
  (\bibinfo{collaboration}{SANE Collaboration}), \bibinfo{journal}{{to be
  published}}  (\bibinfo{year}{{2017}}).

\bibitem[{\citenamefont{Leemann et~al.}(2001)\citenamefont{Leemann, Douglas,
  and Krafft}}]{100}
\bibinfo{author}{\bibfnamefont{C.~W.} \bibnamefont{Leemann}},
  \bibinfo{author}{\bibfnamefont{D.~R.} \bibnamefont{Douglas}},
  \bibnamefont{and} \bibinfo{author}{\bibfnamefont{G.~A.}
  \bibnamefont{Krafft}}, \bibinfo{journal}{Annual Review of Nuclear and
  Particle Science} \textbf{\bibinfo{volume}{51}}, \bibinfo{pages}{413}
  (\bibinfo{year}{2001}).

\bibitem[{12G(Last updated {2016})}]{12GeV}
 (\bibinfo{year}{Last updated {2016}}),
  \urlprefix\url{https://www.jlab.org/12GeV/}.

\bibitem[{\citenamefont{{M. Hauger \emph{et al.}}}(2001)}]{104}
\bibinfo{author}{\bibnamefont{{M. Hauger \emph{et al.}}}},
  \bibinfo{journal}{Nucl. Instrum. Meth.} \textbf{\bibinfo{volume}{A462}},
  \bibinfo{pages}{382} (\bibinfo{year}{2001}), \eprint{nucl-ex/9910013},
  \urlprefix\url{http://www.sciencedirect.com/science/article/pii/S0168900201001978}.

\bibitem[{\citenamefont{{C. Yan \emph{et al.}}}(1995)}]{101}
\bibinfo{author}{\bibnamefont{{C. Yan \emph{et al.}}}}, \bibinfo{journal}{Nucl.
  Instrum. Meth.} \textbf{\bibinfo{volume}{A365}}, \bibinfo{pages}{46}
  (\bibinfo{year}{1995}),
  \urlprefix\url{http://www.sciencedirect.com/science/article/pii/0168900295005048}.

\bibitem[{\citenamefont{Yan et~al.}(2005)\citenamefont{Yan, Sinkine, and
  Wojcik}}]{102}
\bibinfo{author}{\bibfnamefont{C.}~\bibnamefont{Yan}},
  \bibinfo{author}{\bibfnamefont{N.}~\bibnamefont{Sinkine}}, \bibnamefont{and}
  \bibinfo{author}{\bibfnamefont{R.}~\bibnamefont{Wojcik}},
  \bibinfo{journal}{Nucl. Instrum. Meth.} \textbf{\bibinfo{volume}{A539}},
  \bibinfo{pages}{1} (\bibinfo{year}{2005}).

\bibitem[{\citenamefont{{J. D. Maxwell \emph{et al.}}}(2018)}]{SANENIM}
\bibinfo{author}{\bibnamefont{{J. D. Maxwell \emph{et al.}}}},
  \bibinfo{journal}{Nucl. Instrum. Meth.} \textbf{\bibinfo{volume}{A885}},
  \bibinfo{pages}{145} (\bibinfo{year}{2018}), \eprint{1711.09089}.

\bibitem[{\citenamefont{{T. D. Averett \emph{et al.}}}(1999)}]{136}
\bibinfo{author}{\bibnamefont{{T. D. Averett \emph{et al.}}}},
  \bibinfo{journal}{Nucl. Instrum. Meth.} \textbf{\bibinfo{volume}{A427}},
  \bibinfo{pages}{440} (\bibinfo{year}{1999}).

\bibitem[{\citenamefont{Crabb and Meyer.}(1997)}]{137}
\bibinfo{author}{\bibfnamefont{D.}~\bibnamefont{Crabb}} \bibnamefont{and}
  \bibinfo{author}{\bibfnamefont{W.}~\bibnamefont{Meyer.}},
  \bibinfo{journal}{Ann. Rev. Nucl. Part. Sci.} \textbf{\bibinfo{volume}{47}},
  \bibinfo{pages}{67} (\bibinfo{year}{1997}),
  \urlprefix\url{http://arjournals.annualreviews.org/doi/abs/10.1146/annurev.nucl.47.1.67}.

\bibitem[{\citenamefont{Pierce et~al.}(2014)\citenamefont{Pierce, Maxwell, and
  Keith}}]{138}
\bibinfo{author}{\bibfnamefont{J.}~\bibnamefont{Pierce}},
  \bibinfo{author}{\bibfnamefont{J.}~\bibnamefont{Maxwell}}, \bibnamefont{and}
  \bibinfo{author}{\bibfnamefont{C.}~\bibnamefont{Keith}},
  \bibinfo{journal}{Nucl. Instrum. Meth.} \textbf{\bibinfo{volume}{A738}},
  \bibinfo{pages}{54} (\bibinfo{year}{2014}),
  \urlprefix\url{http://www.sciencedirect.com/science/article/pii/S0168900213016999}.

\bibitem[{\citenamefont{Berz}(1995)}]{147}
\bibinfo{author}{\bibfnamefont{M.}~\bibnamefont{Berz}},
  \bibinfo{journal}{Technical Report, Michigan State University}
  (\bibinfo{year}{1995}).

\bibitem[{\citenamefont{Bosted and Christy}(2008)}]{Bosted:2007xd}
\bibinfo{author}{\bibfnamefont{P.~E.} \bibnamefont{Bosted}} \bibnamefont{and}
  \bibinfo{author}{\bibfnamefont{M.~E.} \bibnamefont{Christy}},
  \bibinfo{journal}{Phys. Rev.} \textbf{\bibinfo{volume}{C77}},
  \bibinfo{pages}{065206} (\bibinfo{year}{2008}), \eprint{0711.0159}.

\bibitem[{\citenamefont{Bosted}(1995)}]{Bosted:1994tm}
\bibinfo{author}{\bibfnamefont{P.~E.} \bibnamefont{Bosted}},
  \bibinfo{journal}{Phys. Rev.} \textbf{\bibinfo{volume}{C51}},
  \bibinfo{pages}{409} (\bibinfo{year}{1995}).

\end{thebibliography}

\end{document}